\documentclass[aps,prx,twocolumn,longbibliography,showpacs,amsmath,amssymb,amsfonts,superscriptaddress,nofootinbib]{revtex4-1}
\usepackage{graphicx}
\usepackage{dcolumn}
\usepackage{bm}
\usepackage[usenames,dvipsnames]{color}
\usepackage{multirow}
\usepackage{hyperref}
\usepackage{soul}
\usepackage{natbib}
\usepackage{amsmath, amssymb}
\usepackage{bbold}
\usepackage{physics}
\usepackage{cancel}

\definecolor{magenta}{rgb}{0.8,0.2,0.8}

\begin{document}



\title{Nonequilibrium criticality driven by Kardar-Parisi-Zhang fluctuations in the synchronization of oscillator lattices}

\author{Ricardo Guti\'errez }
\author{Rodolfo Cuerno}
\affiliation{Grupo Interdisciplinar de Sistemas Complejos (GISC), Departamento de Matemáticas, Universidad Carlos III de Madrid, 28911 Legan{\'e}s, Madrid, Spain}

\begin{abstract}
The synchronization of oscillator
ensembles is pervasive throughout nonlinear science, from classical or quantum mechanics to biology, to human assemblies. Traditionally, the main focus has been the identification of threshold parameter values for the transition to synchronization as well as the nature of such transition.
Here, we show that considering an oscillator lattice as a discrete growing interface provides unique insights into the dynamical process whereby the lattice reaches synchronization for long times. Working on a generalization of the celebrated Kuramoto model that allows for odd or non-odd couplings, we elucidate synchronization of oscillator lattices as an instance of generic scale invariance, whereby the system displays space-time criticality, largely irrespective of parameter values. The critical properties of the system (like scaling exponent values and the dynamic scaling Ansatz which is satisfied) happen to fall into universality classes of kinetically rough interfaces with columnar disorder, namely, those of the Edwards-Wilkinson (equivalently, the Larkin model of an elastic interface in a random medium) or the Kardar-Parisi-Zhang (KPZ) equations, for Kuramoto (odd) coupling
and
generic (non-odd) couplings, respectively.
From the point of view of kinetic roughening, the critical properties we find turn out to be quite innovative, especially concerning the statistics of the fluctuations as characterized by their probability distribution function (PDF) and covariance. While the latter happens to be that of the Larkin model irrespective of the symmetry of the coupling, in the generic non-odd coupling case the PDF turns out to be the Tracy-Widom distribution associated with the KPZ nonlinearity. This brings the synchronization of oscillator lattices into a remarkably large class of strongly-correlated, low-dimensional (classical and quantum) systems with strong universal fluctuations.

\end{abstract}



\maketitle

\section{Introduction}

From chirping crickets to Josephson junctions and quantum oscillators, passing through cells in the heart and in the brain, a huge variety of systems across all of science exhibit synchronous dynamics \cite{pikovsky,walter}. While the scientific study of synchronization can be traced back in time to the work of Christiaan Huygens in the 17th century, it is in the last few decades that it has become a central concept in nonlinear and complex dynamical systems, as a pervasive form of emerging collective dynamics. It is frequently studied using models of phase oscillators (i.e.\! idealized limit-cycle oscillators), as in the well-known Kuramoto model \cite{kuramoto_book,acebron}, though in fact it has been studied in low-dimensional chaotic systems as well \cite{boccaletti}, even in complex networks of such systems \cite{arenas}.

Another seemingly unrelated subject that has focused a great deal of attention in the last few decades is the study of surface kinetic roughening \cite{halpinhealy,krug97}, which also unifies a great diversity of nonequilibrium phenomena, from the production of thin solid films to the growth of bacterial colonies, or the formation of coffee rings \cite{barabasi}. The fine details of many experimental systems and theoretical models in this context have been found to become irrelevant at sufficiently large space and time scales, where they show traits of universality akin to that of equilibrium critical dynamics \cite{goldenfeld,tauber14}. Crucially, however, now space-time criticality does not require adjusting control parameters to precise critical values but appears, rather, over a region of parameter space with nonzero measure. Thus, surface kinetic roughening constitutes an important instance of generic scale invariance (GSI) \cite{grinstein91,grinstein95,belitz05}, a concept which is in turn closely related to that of self-organized criticality \cite{dickman00,pruessner12}.


A key player in the GSI realm is the Kardar-Parisi-Zhang (KPZ) stochastic equation \cite{kardar,kriecherbauer10,takeuchi}, whose universality class is recently proving paradigmatic for the space-time critical behavior of fluctuations in low-dimensional, strongly correlated systems. Examples range from non-quantum systems like active matter \cite{chen16}, turbulent liquid crystals \cite{takeuchi11}, stochastic hydrodynamics \cite{mendl13}, colloidal aggregation \cite{yunker13}, thin-film deposition \cite{orrillo17}, reaction-diffusion processes \cite{nesic14}, or random geometry \cite{santalla15}, to the quantum realm, including exciton polariton condensates \cite{altman15,fontaine} or quantum entanglement \cite{nahum17}, integrable and non-integrable quantum spin chains \cite{gopalakrishnan19,ljubotina19,denardis21,wei22}, or electronic fluids \cite{protopopov21}. Probably, such a ubiquity for the KPZ universality is in turn related with the fact that the statistics of fluctuations is described \cite{kriecherbauer10,takeuchi} by the Tracy-Widom (TW) family of probability density functions (PDF) for the maximum eigenvalue of Hermitian random matrices \cite{tracy09,Fortin2015}. Indeed, TW statistics is recently being found ubiquitously \cite{makey20} across length scales in natural and technological systems, providing an analog of Gaussian statistics for correlated variables.

Connections between oscillator synchronization and surface kinetic roughening have been occasionally pointed out in the literature ---see e.g.\! Refs.\ \cite{sakaguchi,grinstein,pikovsky} and others therein---, often through mappings to the KPZ equation, to its Gaussian approximation, the so-called Edwards-Wilkinson (EW) equation \cite{edwards,barabasi}, or variations thereof. For instance, a discretized KPZ equation has been recently shown to approximate a noisy Kuramoto-Sakaguchi model of an (one or two-dimensional) oscillator lattice \cite{lauter}, while the related compact KPZ equation has been shown to describe the dynamics of the phase of a driven-dissipative Bose-Einstein condensate of exciton polaritons \cite{sieberer}. Even more recently, randomness in the natural frequencies of the latter system has been shown \cite{moroney} not to destroy synchronization and KPZ scaling has been experimentally reported in a 1D polariton condensate \cite{fontaine}. In a related context, KPZ universality is known to be relevant in the hydrodynamic phase fluctuations of spatially-extended nonequilibrium oscillating systems, see e.g.\! Refs.\ \cite{chate1995,manneville1996}. However, a systematic assessment of the relevance and implications of KPZ fluctuations to the synchronization of oscillator lattices does not seem available in the literature yet.

In this paper, we aim to fill in this gap. Focusing on a generalization of the model of one-dimensional (1D) oscillator lattices addressed in Refs.\ \cite{lauter,moroney} ---which allows for both odd (as in the seminal Kuramoto model \cite{kuramoto1984,kuramoto_book}) and non-odd coupling functions among oscillators---, we unambiguously elucidate synchronization in these systems as an instance of GSI, with anomalous forms of scaling and universal fluctuation statistics 
which are most conveniently phrased in terms of those characterizing GSI in surface kinetic roughening, and in which KPZ fluctuations play an important role. The lack of awareness that the dynamics of synchronizing oscillators is endowed with universal features due to nonequilibrium criticality probably explains why the dynamical process whereby synchronization is achieved has remained poorly studied.

Previous analytical results \cite{strogatz} show that synchronization of oscillator lattices is not possible in the thermodynamic limit for odd coupling functions. Viewing the oscillator array as an evolving interface, we show this lack of synchronization to be a consequence of so-called super-rough kinetic roughening \cite{dassarma}, whereby the local slopes along the interface only stabilize upon saturation \cite{schroeder93,lopez}. Moreover, the odd symmetry of the coupling function becomes the condition for which synchronization features the precise scaling behavior of the EW model with so-called columnar noise, known as the Larkin model in the context of elastic interfaces in disordered media; see, e.g., Ref.~\cite{purrello} and references therein. For generic coupling functions, on the other hand, we show that synchronization becomes possible for arbitrarily large systems \cite{ostborn}, leading to the growth of faceted interfaces (oscillator clusters) where the slopes stabilize earlier \cite{ramasco}. The large-scale effective description is now given by the KPZ equation with columnar noise,
which however is not in the universality class of the celebrated KPZ equation \cite{szendro}. Different synchronous dynamics thus map into different surface growth models.

From the stand point of kinetic roughening, the dynamic scaling Ansatz satisfied by the space-time fluctuations of the oscillator array is anomalous \cite{schroeder93,dassarma,lopez,ramasco}, akin to those found in many systems with morphological instabilities and/or quenched disorder \cite{cuerno04}. What is more striking is that for non-odd coupling functions, phase fluctuations follow the ubiquitous TW PDF, one of the hallmarks of KPZ universality, in spite of the fact that neither the scaling Ansatz, nor the critical exponent values, nor the covariance of the fluctuations are those of the 1D KPZ universality class. This is in line with very recent observations in continuous \cite{rodriguez-fernandez21} and discrete \cite{marcos22} models that, somewhat unexpectedly, the fluctuation PDF and covariance, and the scaling exponents are all independent traits of a GSI universality class.

This paper is organized as follows. In Section II we describe first the connection between synchronization and kinetic roughening, followed by a discussion of the observables of interest. Section III contains a description of the particular model to be considered in the numerics and a general discussion and results on the synchronous dynamics. Section IV is devoted to the study of synchronization with the Kuramoto (sine) coupling form, for which we solve the continuum-limit (linear) Larkin model. Numerical simulations reveal a super-rough scaling adequately described by the analytical results. In Section V we study synchronization with a non-odd coupling function, whose effective description yields a (nonlinear) KPZ equation with columnar noise. Sections IV and V also contain the study of the one- and two-point statistics of fluctuations that yields Gaussian (respectively, TW) statistics for odd (respectively, non-odd) couplings, but the same covariance of the Larkin model in all cases. Finally, in Section VI we provide some concluding remarks and ideas for future work. Additional numerical results are organized into five appendices at the end.

\section{Synchronization vs Kinetic Roughening}

\subsection{General lattices of phase oscillators}

We consider a system of $L^d$  {\it phase oscillators} at the sites of a $d$-dimensional hypercubic lattice of linear size $L$. Each oscillator is an idealized dissipative dynamical system with an attracting limit cycle, whose state is given by a phase $\phi_i(t)$. The time evolution is determined by its {\it intrinsic frequency} $\omega_i$ and the interactions with its neighbors through a smooth, $2\pi$-periodic {\it coupling function} $\Gamma(\phi_j -\phi_i)$,
\begin{equation}
\frac{d \phi_i}{d t} = \omega_i + \sum_{j\in \Lambda_i} \Gamma(\phi_j -\phi_i),\ i=1,2,\ldots, L^d,
\label{eq1}
\end{equation}
where $\Lambda_i$ is the set of $2 d$ neighbors of site $i$. The intrinsic frequencies are independent and identically distributed according to a probability density $g(\omega)$ with zero mean and finite variance, i.e.\! $\langle \omega_i \rangle = \int d\omega g(\omega)\,  \omega = 0$  and $\langle \omega_i \omega_j \rangle = \delta_{ij}\int d\omega g(\omega)\, \omega^2=  2 \sigma \delta_{i j}$, where $\delta_{i j}$ is the Kronecker delta. A nonzero mean would only introduce a uniform frequency shift, easily removable by moving to the rotating frame. Moreover, we assume that the distribution is even, $g(-\omega) = g(\omega)$. The coupling function $\Gamma(\phi_j -\phi_i)$ is assumed to include a {\it coupling strength} $K$, i.e.\! an overall nonnegative proportionality factor with dimensions of inverse time.

The {\it effective frequencies} of oscillation are defined as
\begin{equation}
\omega^\text{eff}_i \equiv \lim_{T\to\infty}\frac{\phi_i(\tau+T)-\phi_i(\tau)}{T},\ i=1,2,\ldots, L^d,
\label{omegaeff}
\end{equation}
where $[0,\tau]$ is a time interval sufficiently long to contain the transient behavior, and the limit is assumed to exist. Oscillators that evolve at the same effective frequency are said to be {\it frequency entrained} or {\it frequency locked}. The kinetic-roughening observables that we consider focus on this (time-averaged) form of synchronization, which does not require the {\it instantaneous frequencies} $d \phi_i/d t$ to become strictly identical at all times.

\subsection{Continuum approximation}
\label{contlim}

Generalizing the 1D approach of Ref.~\cite{kuramoto1984}, we write the positions of the oscillators as vectors in continuous space, ${\bf x} = (x_1, x_2, \ldots, x_d)\in \mathbb{R}^d$, so the phase of oscillator $i$, $\phi_i$, is now denoted $\phi({\bf x})$, and the neighboring oscillators are placed at positions  ${\bf x}\pm a {\bf e}_k$, where ${\bf e}_k$ for $k=1,2,\ldots, d$ is a canonical basis vector. Thus Eq.~(\ref{eq1}) becomes
\begin{align}
\partial_t \phi({\bf x},t) = \omega({\bf x}) &+ \sum_{k=1}^d \left[\Gamma(\phi({\bf x}+ a {\bf e}_k,t)  -\phi({\bf x},t))\right.\nonumber\\
&\left.+ \Gamma(\phi({\bf x}- a {\bf e}_k,t)  -\phi({\bf x},t))\right],
\label{eq2}
\end{align}
where $\langle \omega({\bf x}) \rangle = 0$, and $\langle \omega({\bf x})\omega({\bf x}') \rangle = 2 \sigma \delta({\bf x} -  {\bf x}')$, using the Dirac delta. We will focus on a coarse-grained description where $a$ is assumed to be small compared to the wavelengths over which the phase field $\phi({\bf x},t)$ fluctuates.  Taylor-expanding the phase field around ${\bf x}$,
\begin{align}
\phi({\bf x}\pm a {\bf e}_k,t) - \phi({\bf x}) &= \pm a \partial_k \phi({\bf x},t) + \frac{1}{2} a^2  \partial_k^2\phi({\bf x},t)\nonumber\\
&\pm \frac{1}{6} a^3  \partial_k^3\phi({\bf x},t) + \mathcal{O}(a^4),
\end{align}
where $\partial_k$ is shorthand for the partial derivative with respect to the $k$-th coordinate $x_k$. In Eq.~(\ref{eq2}), after Taylor-expanding the coupling functions, terms of odd order in $a$ vanish due to the $a\to -a$ symmetry of the coupling term, yielding
\begin{align}
\partial_t \phi ({\bf x},t)  &=\omega({\bf x}) + 2 d \Gamma(0)
+ a^2 \Gamma^{(1)}(0)\!\sum_{k=1}^d\!\partial_k^2 \phi({\bf x},t) \nonumber\\
&+  a^2 \Gamma^{(2)}(0)\!\sum_{k=1}^d\!(\partial_k \phi({\bf x},t) )^2 +  \mathcal{O}(a^4),
\label{eq3}
\end{align}
where $\Gamma^{(n)}$ denotes the $n$-th derivative of $\Gamma$.

For a relatively slow spatial variation of the phase field $\phi({\bf x},t)$, as occurs for coupling strengths $K$ well into the synchronized regime, it may be reasonable to neglect terms of order higher than $a^2$, which is the dominant one for spatial coupling in the oscillating medium. By analogy with surface growth, we will refer to this as a {\it small-slope approximation} which yields the effective continuum equation
\begin{equation}
\partial_t \phi({\bf x},t) =   \omega^*({\bf x}) + \nu \nabla^2 \phi({\bf x},t) + \frac{\lambda}{2} [\nabla \phi({\bf x},t)]^2,
\label{eq4}
\end{equation}
where $\omega^*({\bf x})\equiv  \omega({\bf x}) + 2 d \Gamma(0)$, and as usual the Laplacian $\nabla^2 \phi({\bf x},t) \equiv \sum_{k=1}^d\!\partial_k^2 \phi({\bf x},t)$ and the squared norm of the gradient  $[\nabla \phi({\bf x},t)]^2\equiv \sum_{k=1}^d\![\partial_k \phi({\bf x},t)]^2$.  The appropriateness of the truncation for specific cases will be discussed when making the comparison between predictions derived from it and results  based on the direct numerical integration of particular instances of Eq.~(\ref{eq1}). 
A similar continuum approximation for extended systems with time-delayed couplings was analyzed and applied to the study of long-wavelength modes of the vertebrate segmentation clock in Ref.\! \cite{ares}.


Equation \eqref{eq4} features the same deterministic derivative terms as the KPZ equation \cite{kardar}. We have introduced two parameters, $\nu \equiv  a^2 \Gamma^{(1)}(0)$ and $\lambda/2 \equiv  a^2 \Gamma^{(2)}(0)$, following the standard notation in the surface growth literature, where they quantify smoothening surface tension and interface growth along the local surface normal direction, respectively \cite{barabasi}. In the case of the oscillators such names at most provide an intuitive meaning to the parameters, which are not necessarily positive. This formal connection between oscillator lattices and rough interfaces has been pointed out on several occasions in the literature, with some recent works even exploiting it for the study of synchronization in novel scenarios \cite{lauter,moroney}.

Notice, however, that in Eq.~\eqref{eq4} the noise term, $\omega^*({\bf x})$ is time-independent, in contrast with the time-dependent noise of the standard KPZ equation. In the presence of such quenched disorder the system evolves \emph{deterministically} from the initial condition. The dynamics is akin to that of a growth process in a medium for which the disorder values depend on the substrate coordinate ${\bf x}$ but not on the local value of the interface ``height'' $\phi({\bf x},t)$. In the kinetic roughening literature, Eq.\ \eqref{eq4} is  known as the KPZ equation with \emph{columnar} noise \cite{halpinhealy, szendro}. Its linear version, obtained for $\lambda = 0$, is the EW equation with columnar noise, known also as the Larkin model.

\subsection{Morphological analysis}
\label{obs}

In our analysis we consider the phase field $\phi({\bf x},t)$ as if it were describing the height $h({\bf x},t)$ of an interface growing above point ${\bf x}\in \mathbb{R}^d$ on a $d$-dimensional substrate, at time $t$. In surface kinetic roughening processes
\cite{barabasi,halpinhealy,krug97}, the fluctuations of the local height around the mean value are captured by the {\it global width} or {\it roughness}
\begin{equation}
W(L,t) \equiv \langle \overline{[h({\bf x},t)-\overline{h}]^2} \rangle^{1/2},
\end{equation}
where the overbar denotes a spatial average in a system of linear (substrate) size $L$ and the angular brackets denote averaging over different noise realizations.  GSI implies that surface height values are statistically correlated for distances smaller than a correlation length $\xi(t)$ which increases with time as a power law, $\xi(t) \sim t^{1/z}$, where $z$ is the so-called {\it dynamic exponent}. Such an increase takes place until $\xi(t)$ reaches a value comparable to $L$, which results in the width saturating at a steady-state, size-dependent value $W(L,t\gg L^z) \sim L^\alpha$. Here, $\alpha$ is the so-called {\it roughness exponent}, which is related with the fractal dimension of the interface profile $h(\mathbf{x},t)$. In a wide variety of physical contexts and conditions, the global roughness satisfies the {\it Family-Vicsek (FV) scaling Ansatz}
\cite{barabasi,halpinhealy,krug97,vicsek}
\begin{equation}
W(L,t) = t^{\beta} f(L/\xi(t)),
\label{fv}
\end{equation}
where the scaling function $f(y) \sim y^\alpha$ for $y\ll 1$, while $f(y)$ reaches a constant value for $y\gg 1$. The ratio $\beta = \alpha/z$ is known as the {\it growth exponent}, and characterizes the short-time behavior of the roughness, $W(t)\sim t^{\beta}$. Notably, the FV Ansatz is verified by classical models of equilibrium critical dynamics \cite{tauber14}. Away from equilibrium, it is also verified by representatives of important universality classes of kinetic roughening, like those of the KPZ and EW equations, which are characterized by the set of $(\alpha,z)$ exponent values and their dependence on the substrate dimension $d$ \cite{barabasi,halpinhealy,krug97}. In this sense, kinetic roughening extends classical equilibrium critical dynamics far from equilibrium \cite{tauber14}.

Beyond global quantities like $W(t)$, the GSI occurring in kinetic roughening systems also impacts the behavior of correlation functions. A particularly useful one is the {\it height-difference correlation function},
\begin{equation}
    G({\bf r},t) \equiv \langle \overline{[h({\bf x}+{\bf r},t) - h({\bf x},t)]^2} \rangle.
\label{grt}
\end{equation}
In our cases of interest, due to rotational invariance, the correlations only depend on $\ell \equiv |{\bf r}|$. For FV scaling 
$G(\ell,t)$ scales differently depending on how $\ell$ compares with the correlation length \cite{barabasi,halpinhealy,krug97},
\begin{equation}
    G(\ell,t) \sim \left\{
    \begin{array}{lr}
        t^{2 \beta},& \text{if } t^{1/z} \ll \ell, \\
        \ell^{2 \alpha}, & \text{if } \ell  \ll t^{1/z}.
    \end{array}
    \right\} = \ell^{2\alpha} g(\ell/\xi(t)) ,
\label{glt}
\end{equation}
where we are assuming $\ell < L$ and $g(y)$ is a suitable scaling function \cite{barabasi,krug97}. In fact, $G(\ell,t)$ scales like the square of the {\it local width} $w(\ell,t) \equiv \langle \overline{[h({\bf x},t)-\overline{h}]^2} \rangle^{1/2}$ (the spatial average is here restricted to a region of linear size $\ell$), $G(\ell,t) \sim w^2(\ell,t)$. One can see that the scaling form in Eq.\ (\ref{glt}) only reflects the growth and saturation discussed above for the whole system, but now restricted to a region of linear size $\ell$ \cite{lopezphysa}. A related correlation function that is also frequently studied \cite{takeuchi} is the so-called {\it height covariance}
\begin{equation}
C(\mathbf{r},t) \equiv \langle \overline{h(\mathbf{x},t) h(\mathbf{x}+\mathbf{r},t)} \rangle - \langle \bar{h}(t) \rangle^2 ,
    \label{eq:cov}
\end{equation}
such that, again under the assumption of rotational invariance, $G(\ell,t)= 2 [W^2(t)-C(\ell,t)]$ \cite{krug97}.


Whenever the roughness exponent $\alpha \geq 1$, it is useful to consider \cite{lopezphysa, siegert96} an alternative two-point correlation function, namely, the surface {\it structure factor}, i.e.\! the power spectral density of the height fluctuations, defined as
\begin{equation}
S({\bf k},t) \equiv \langle \hat{h}({\bf k}, t)\hat{h}(-{\bf k}, t)\rangle = \langle |\hat{h}({\bf k}, t)|^2\rangle,
\end{equation}
where $\hat{h}({\bf k},t)\equiv\mathcal{F}[h(\mathbf{x},t)]$ is the space Fourier transform of $h(\mathbf{x},t)$ and $\mathbf{k}$ is $d$-dimensional wave vector. The FV Ansatz now reads
\begin{equation}
S({\bf k},t) = k^{-(2  \alpha +d)} s_{\rm FV}(k t^{1/z}) ,
\label{SkFV}
\end{equation}
with $s_{\rm FV}(y)$ approaching a constant value for $y\gg 1$ and $s_{\rm FV}(y) \propto y^{2\alpha +d}$ for $y\ll 1$. This can be derived by realizing that $W^2(L,t)$ equals the integral of $S({\bf k},t)$ over wave vector space (Parseval's theorem) \cite{barabasi}. Likewise, $S(\mathbf{k},t)$ is analytically related with, e.g., $G(\mathbf{r},t)$ via space Fourier transforms \cite{krug97}.

In the case of a system of oscillators, analogous observables to the global roughness and the correlation functions, to be denoted as $W_\phi(L,t)$, $G_\phi({\bf r},t)$, $C_\phi({\bf r},t)$ and $S_\phi({\bf k},t)$,
 are simply defined by replacing the height field $h({\bf x},t)$ by the phase field $\phi({\bf x},t)$. They will be the main objects of our analysis in the sections to follow. Regarding two-point correlations, we will be particularly interested in their value at a distance of one site, $G_\phi(\ell = 1,t)$, which will be referred to as the {\it average squared slope}, and denoted as  $\langle\overline{(\Delta \phi)^2} \rangle$. Our interest lies in finite systems, where saturation may be eventually attained. The key point is that differences between oscillator phases that do not evolve at the same effective frequency $\omega^{\text{eff}}$ must grow steadily in time for long times. Thus the phases of two oscillators with effective frequencies $\omega^{\text{eff}}_1$ and $\omega^{\text{eff}}_2$ eventually separate, $\phi_2(t) - \phi_1(t) \sim (\omega^{\text{eff}}_2-\omega^{\text{eff}}_1)t$. For this reason, the saturation of $W_\phi(L,t)$ [or equivalently, that of $S_\phi({\bf k},t)$ or $G_\phi({\bf r},t)$] as $t\to\infty$, which shows that the phase differences stop growing at some time, indicates the presence of synchronization in the sense mentioned above.

\subsection{Anomalous scaling and universal fluctuations in the synchronization process}
\label{anom}

Two further aspects of surface growth, which have been the focus of much recent research, will turn out to be indispensable for the analysis of the synchronization problem. They both challenge the traditional characterization of kinetic-roughening universality classes in terms of just two independent exponents appearing in the FV dynamic scaling Ansatz, Eq.\ (\ref{fv}). One is the existence of growth processes with {\it anomalous scaling} properties \cite{schroeder93,dassarma,krug97,lopez,ramasco,cuerno04}. While for standard FV systems height fluctuations at local distances $\ell \ll L$ scale with the same roughness exponent as global fluctuations do at distances comparable with the system size $L$, in systems displaying anomalous scaling local and global fluctuations scale with different roughness exponents, i.e. $w(\ell,t\gg \ell^z) \sim \ell^{\alpha_\text{loc}}$ with $\alpha_\text{loc} \neq \alpha$.

The anomalous scaling that occurs in our work  is most conveniently identified by means of the structure factor, as a new independent exponent $\alpha_s$ appears (termed {\it spectral roughness exponent}), in the dominant contribution in Fourier space, namely \cite{ramasco},
\begin{equation}
S(k,t) = k^{-(2  \alpha +d)} s(k t^{1/z}) ,
\label{Sk}
\end{equation}
where $s(y) \propto y^{2(\alpha - \alpha_s)}$ for $y\gg 1$ and $s(y) \propto y^{2\alpha +d}$ for $y\ll 1$.
Equation \eqref{Sk} generalizes the FV Ansatz, Eq.\ \eqref{SkFV}, which is retrieved if $\alpha_s=\alpha$. In general, the value of $\alpha_s$ and its relation with respect to $\alpha$ determines the type of anomalous scaling \cite{ramasco}.

We will be interested in cases such that $\alpha_s>1$, for which the correlation function scales as \cite{lopezphysa}
\begin{equation}
    G(\ell,t) \sim \left\{
    \begin{array}{lr}
        t^{2 \beta},& \text{if } t^{1/z} \ll \ell \ll L, \\
        \ell^{2 \alpha_\text{loc}} t^{2(\alpha-\alpha_\text{loc})/z}, & \text{if } \ell  \ll t^{1/z}  \ll L.
    \end{array}
\right.
\label{glt_mod}
\end{equation}
This means that the two-point correlations keep increasing (anomalously) with time even at distances which are smaller than the correlation length, at which they saturate in case of FV scaling; in contrast, now they only saturate at $\ell^{2 \alpha_\text{loc}} L^{2(\alpha-\alpha_\text{loc})}$ when $t^{1/z} \sim L$. If $\alpha = \alpha_s >1$, the anomalous scaling is termed super-rough \cite{lopez}, due to the large interface fluctuations that occur. The scaling Ansatz satisfied by the structure factor is FV in this case, but $\alpha_\text{loc} = 1 \neq \alpha$. Otherwise, if $\alpha \neq \alpha_s$ with both exponents being larger than 1, faceted anomalous scaling takes place \cite{ramasco}, and again $\alpha_\text{loc} = 1$. There exist cases (like that of the tensionless KPZ equation \cite{rodriguez-fernandez22}) in which $\alpha_s<1$, leading to so-called intrinsic anomalous scaling for which $\alpha_\text{loc}$ need not equal 1 \cite{lopez}.

A second aspect of surface kinetic roughening that is relevant to our analysis has to do with the {\it statistics of fluctuations}. For rough surfaces, the observable of interest is the PDF of the fluctuations of the heights $h({\bf x},t)$ [in our case, that of the phases $\phi({\bf x},t)$], around their mean.
By a straightforward adaptation of the definition employed in the kinetic roughening literature \cite{kriecherbauer10,halpinhealy,takeuchi}, we will focus on the PDF of
\begin{equation}
\varphi_i\equiv \frac{\delta \phi_i(t_0+\Delta t) - \delta \phi_i(t_0)}{(\Delta t)^\beta} ,
\label{fluct}
\end{equation}
where $\delta \phi_i(t) = \phi_i(t) - \overline{\phi}(t)$, $t_0$ is a reference time beyond the initial transient dynamics, and $t_0 +\Delta t$ is some intermediate time within the growth regime. The division by $(\Delta t)^\beta$ removes the systematic increase of the fluctuations in time so that, remarkably, the PDF of $\varphi_i$ reaches a universal, time-independent form \cite{kriecherbauer10,halpinhealy,takeuchi}.
Important examples in the kinetic roughening literature are e.g.\ the Gaussian distribution for the linear EW equation \cite{barabasi,krug97} and a TW PDF (whose precise form depends, e.g., on boundary conditions) for the KPZ equation \cite{kriecherbauer10,halpinhealy,takeuchi}.

Our main numerical findings can be summarized as follows: 1) for odd (Kuramoto) coupling the scaling is super-rough, the exponents and scaling Ansatz are those of the EW equation with columnar noise (Larkin model), and the fluctuations follow a Gaussian PDF; 2) for generic couplings the scaling is faceted, the exponents and the scaling Asantz are those of the KPZ equation with columnar noise, and the fluctuations follow a TW PDF. The covariance (\ref{eq:cov}), however, is that of the Larkin model in all cases. Thus, the case of non-odd coupling seems to be the first known example of kinetic roughening displaying TW statistics but not an Airy covariance. These results are based on a 1D model of phase oscillators described and explored in Sec.~\ref{prelim}, whose scaling and fluctuations are studied in Secs.~\ref{odd} (for odd coupling) and \ref{nonodd} (for non-odd coupling).
\begin{center}
\begin{table}
\renewcommand{\arraystretch}{1.5}
\begin{tabular}{||c || c|  c||}
\hline\hline
 $\Gamma(\Delta \phi)$ & Odd & Generic (non-odd) \\
 \hline\hline
 Continuum & Columnar EW \cite{purrello} & Columnar KPZ \cite{szendro} \\
 \hline
 Scaling & Super-rough \cite{dassarma}  & Faceted \cite{ramasco}\\
 \hline
 Exponents \ & $\alpha=3/2, z=2$          &  $\alpha\approx 1.07, z\approx 1.39$  \\
  \hline
 Anom.\ Exp.\ & $\alpha_s=3/2, \alpha_\text{loc} = 1$           &   $\alpha_s\approx 1.40, \alpha_\text{loc} \approx 0.96$  \\
  \hline
 PDF & Gaussian & TW \\
 \hline
 $C_\phi({\bf r},t)$ & Larkin model & Larkin model \\
 \hline\hline
\end{tabular}
\caption{{\sf \bf Summary of correspondences between oscillator models and kinetic roughening equations, and main scaling and fluctuation properties.}
Depending on whether the coupling is odd or generic (non-odd) we find different continuum approximations (corresponding to the two main models of kinetic roughening with columnar noise), with different forms of anomalous scaling, exponent values, and fluctuation PDF. The covariances take the same form, however. References are given to works on the corresponding interface equations. Results on the scaling exponents are contained there and confirmed by our simulations for the oscillator models. Results quoted for the PDF and covariance are obtained from our simulations for the oscillator models.
}\label{tab}
\end{table} \end{center}

These findings, together with some more specific details, are summarized in Table \ref{tab}. The exponents are divided into two classes: the standard exponents $\alpha$ and $z$, and the anomalous-scaling exponents $\alpha_s$ and $\alpha_\text{loc}$. In the case of odd coupling, all these exponents are simply read from the exact Larkin model solution (see e.g.\! \cite{purrello}). In the case of generic couplings, they are obtained numerically, and what is presented is a rough approximation, as they slightly change with different coupling parameters. Moreover, they may depend on other nonuniversal details according to what is known on the scaling behavior of the KPZ equation with columnar noise \cite{krug97,szendro,nattermann,krughh}.

\section{Model and Conditions for Synchronization}
\label{prelim}

\subsection{Description of the model}
\label{nummod}

The coupling function $\Gamma(\Delta \phi)$ in Eq.~(\ref{eq1}) is assumed to be $2\pi$-periodic and smooth. One property that has been highlighted in the literature is its symmetry under phase inversion $\Delta \phi \to -\Delta \phi$, i.e.\! whether the function is odd, $\Gamma(\Delta \phi) + \Gamma(-\Delta \phi) = 0$, or not \cite{strogatz, ostborn}. We will see that this symmetry has crucial implications in several dynamical features of synchronization, which can be related to the occurrence of the nonlinearity in the continuum approximation given by Eq.~(\ref{eq4}). An obvious feature already in the oscillator model, Eq.\ \eqref{eq1}, is that only when $\Gamma(\Delta\phi)$ is odd does the system have up-down symmetry, i.e.\! invariance under $\phi_i\to -\phi_i$, in a statistical sense.

Our simulations will be based on oscillator rings, i.e.\! 1D chains with periodic boundary conditions. For simplicity we always assume that the coupling function $\Gamma(\Delta \phi)$ is described by the fundamental term in a Fourier series, so higher harmonics are not considered. Moreover, we assume that $\Gamma(0) = 0$: when an oscillator phase is equal to that of a neighbor, the coupling between them vanishes. Therefore the noise term in the effective dynamics, Eq.\ (\ref{eq4}), is simply given by the intrinsic frequencies, $\omega^*(x)  = \omega(x)$. Under such assumptions, the coupling function can be written, without loss of generality, as
\begin{equation}
\Gamma(\Delta \phi) = K(\sin(\Delta \phi + \delta)-\sin \delta),
\label{gamma}
\end{equation}
for $\delta \in (-\pi,\pi]$. With a slightly different notation, it was previously considered in the seminal Ref.~\cite{sakaguchi}. The odd symmetry $\Gamma(\Delta \phi) + \Gamma(-\Delta \phi) = 0$ is obtained only for $\delta = 0$ and $\pi$, which correspond to $\Gamma(\Delta \phi) = K \sin(\Delta \phi)$ (Kuramoto coupling) and $\Gamma(\Delta \phi) = -K \sin(\Delta \phi)$, respectively.

The model that we study numerically is thus
\begin{equation}
\frac{d\phi_i}{dt} = \omega_i + K[\sin(\phi_{i+1}\!-\!\phi_i\!+\!\delta) +\sin(\phi_{i-1}\!-\!\phi_i\!+\!\delta)-2\sin\delta],
\label{eqsim}
\end{equation}
for $i = 1,2,\ldots, L$, where $\phi_0 \equiv \phi_L$ and $\phi_{L+1} \equiv \phi_1$. The system starts from a homogeneous initial condition, $\phi_i(0)=0$ for all $i$, and is integrated by means of a fourth-order Runge-Kutta algorithm with time step $\delta t = 0.01$. The intrinsic frequencies $\omega_i$ are uniformly distributed over $[-1,1]$, a choice of the distribution $g(\omega)$ that satisfies the properties discussed immediately after Eq.\ (\ref{eq1}). The synchronous dynamics of the system of phase oscillators governed by Eq.\! (\ref{eqsim}) has been explored (except for a uniform shift in the intrinsic frequencies) for the parameter choice $\delta = -\pi/4$ in a recent publication that, motivated by the study of driven-dissipative bosonic systems, examines the transition to synchronization as a diffusion process in a random medium \cite{moroney}. Some of the morphologies to be discussed here present similarities to those displayed in that work, as well as in the study of the growth process arising from the KPZ equation with columnar noise \cite{szendro}.

Another symmetry will be relevant in the following: The coupling function, Eq.\ (\ref{gamma}), changes sign under the simultaneous reversal of the phases and the angle $\delta$. This results into the invariance of Eq.\ (\ref{eqsim}) under $\phi_i \to - \phi_i$ and $\delta \to -\delta$,  in the same statistical sense mentioned above for the invariance under phase reversal (up-down symmetry) of an odd coupling function. In fact the latter can be considered as a particular case of the former invariance for $\delta = 0$ or $\pi$ in the dynamical model, Eq.\ (\ref{eqsim}).

\subsection{Synchronization in 1D systems of phase oscillators: Summary of previous results}
\label{cond}

Strogatz and Mirollo proved \cite{strogatz} that, in order for synchronization to occur in a typical chain of $L$ oscillators, for large $L$, with Kuramoto coupling $\Gamma(\Delta \phi) = \sin (\Delta \phi)$ (and a distribution of intrinsic frequencies with finite mean and variance), a coupling strength of size $\mathcal{O}(\sqrt{L})$ is required, which precludes the existence of synchronization in the thermodynamic limit. Their proof rests on the fact that the sine function is odd. In fact, the (non-)odd character of the coupling function has been pointed out as an important feature for the emergence of synchronization in other works. Thus, Kopell and Ermentrout showed \cite{kopell} that a non-odd coupling leads to frequency entrainment when the intrinsic frequencies satisfy  $|\omega_{i+1} - \omega_i| = \mathcal{O}(1/L)$. For spatially uncorrelated frequencies, Sakaguchi {\it et al.}\! provided evidence \cite{sakaguchi} indicating that a coupling function of the form  $K(\sin(\Delta \phi + \delta)-\sin \delta)$ for $\delta \neq 0$ favors synchronization when compared to the odd $\delta = 0$ case. Their arguments are based on a simplified two-oscillator problem and a mapping of the many-oscillator problem into the Schr\"odinger equation of a particle in a random potential (which presents some similarities with the recent analysis in Ref.~\cite{moroney}), as well as numerical results. A more elaborate theoretical treatment, under some mathematical assumptions concerning the coupling function, was provided by \"Ostborn, who showed \cite{ostborn} that for non-odd couplings synchronous solutions exists and are stable for coupling strengths above a critical value that is independent of the system size. This estimate of the critical coupling strength, together with the one based on the two-oscillator model in Ref.~\cite{sakaguchi}, will be discussed in the next section.

\subsection{Critical coupling strength $K_c$}
\label{kc}

The oscillator model in Eq.~(\ref{eqsim}) has two parameters, namely the angle $\delta$ and the coupling strength $K$. As our interest lies in synchronous dynamics, we will focus on sufficiently strong couplings $K\geq K_c$, where $K_c$ is the critical coupling strength for synchronization, which in general may depend on both $\delta$ and the system size $L$. Further, we will restrict the numerical analysis to parameter choices for which the small-slope approximation leading to Eq.\ (\ref{eq4}) is sensible.

\begin{figure}[t!]
\includegraphics[scale=0.56]{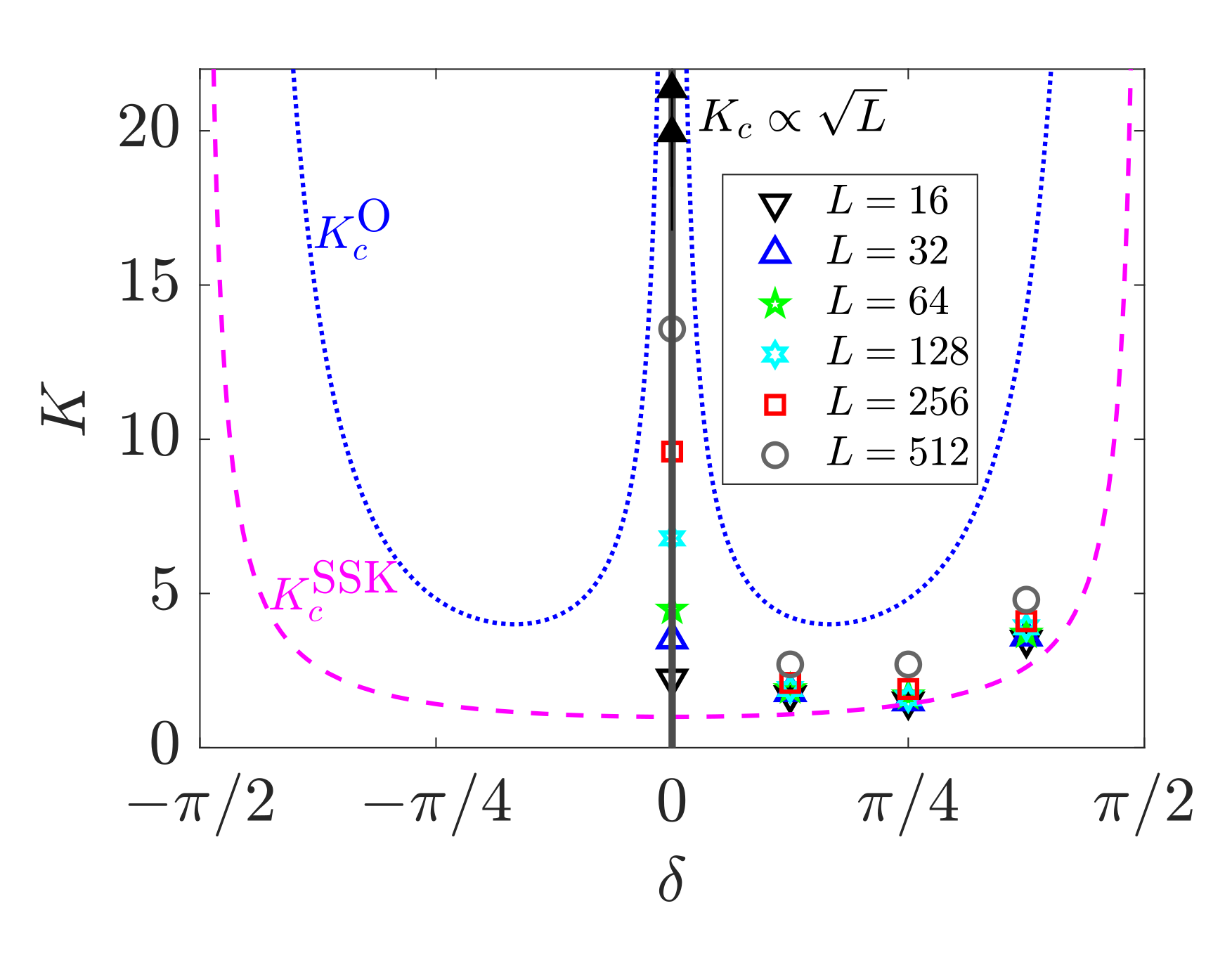}
\vspace{-0.2cm}
\caption{{\sf \bf Theoretical and numerical estimates of the critical coupling strength $K_c$ as a function of $\delta$.}  The estimate based on a two-oscillator model $K_c^\text{SSK}$, Eq.\ (\ref{kssk}), is shown as a magenta dashed line. The many-oscillator estimate $K_c^\text{O}$, Eq.\ (\ref{ko}), is shown as a blue dotted line. Numerical estimates based on the stability of the roughness $W_{\phi}(L,t)$ are represented as isolated points, with symbols depending on the system size $L$ (see legend). It has been proved \cite{strogatz} that synchronization in a typical chain of oscillators for $\delta=0$ requires a critical coupling strength $K_c$ that increases as $\sqrt{L}$, which is also illustrated. See text for definitions and references.}
\label{figPD}
\end{figure}

To gain some insight into the dependence of the critical coupling strength $K_c$ and the stationary slopes $\Delta \phi$ on the parameters, it is useful to consider the simple two-oscillator model introduced in Ref.~\cite{sakaguchi}:
\begin{align}
\frac{d\phi_1}{dt} &= \omega_1 +   K(\sin(\phi_2-\phi_1+ \delta)-\sin \delta),\nonumber\\
\frac{d\phi_2}{dt} &= \omega_2+   K(\sin(\phi_1-\phi_2+ \delta)-\sin \delta).
\label{twoosc}
\end{align}
In terms of the auxiliary variables $\bar{\phi} = (\phi_1 + \phi_2)/2$ and $\Delta \phi = \phi_2 - \phi_1$, and the corresponding intrinsic frequencies $\bar{\omega} = (\omega_1+ \omega_2)/2$ and $\Delta \omega = \omega_2 - \omega_1$, it becomes
\begin{align}
\frac{d\bar{\phi}}{dt} &= \bar{\omega} - K \sin \delta (1-\cos \Delta \phi),\nonumber\\
\frac{d(\Delta \phi)}{dt} &= \Delta \omega - 2 K \cos \delta \sin \Delta \phi.
\label{twooscpre}
\end{align}
Synchronization in this simple model is achieved for $K$ exceeding the critical strength $K_c^\text{SSK} \equiv |\Delta  \omega/2 \cos \delta|$, as only for such coupling can the slope $\Delta \phi$ assume a stationary value. To distinguish this one from the other estimate of the critical coupling strength $K_c$ that we will use, we have added a superscript indicating the initials of the authors in $K_c^\text{SSK}$ \cite{sakaguchi}. Since the distribution of natural frequencies $g(\omega)$ that we consider is uniform over $[-1,1]$, we take as the intrinsic frequency difference in our (many-oscillator) model the upper bound given by $|\Delta \omega| = 2$, which yields
\begin{equation}
 K_c^\text{SSK} = \frac{1}{|\cos \delta|}.
 \label{kssk}
\end{equation}
Its dependence on $\delta$ is shown as a magenta dashed line in Fig.~\ref{figPD}.

As long as $\delta \neq \pm \pi/2$, for $K\geq K_c^\text{SSK}$ the phase difference $\Delta \phi$ stabilizes, and the two-oscillator model achieves synchronization. The stable equilibrium of the second line in Eq.\ (\ref{twooscpre}) is found for $\Delta \phi$ satisfying the following two conditions:
\begin{equation}
\cos \delta \cos \Delta \phi > 0,\ \ \sin \Delta \phi = \frac{\Delta \omega}{ 2 K \cos \delta}.
\label{twoosccond}
\end{equation}
According to the first condition in Eq.\ (\ref{twoosccond}), the stable fixed point is reached for $\Delta \phi \mod 2\pi$ in $(-\pi/2,\pi/2)$ or  in $(-\pi,-\pi/2)\cup (\pi/2,\pi]$, depending on whether $\delta$ is in one or the other subset of $(-\pi,\pi]$. Moreover, the second condition in Eq.\ (\ref{twoosccond}) implies that $|\sin \Delta \phi| = K_c^\text{SSK}/K$. Therefore, as $K (> K_c^\text{SSK})$ increases $\Delta \phi$ gets closer to $0$ or $\pi$ (again, depending on the sign of $\cos \delta$).

\begin{figure*}[t!]
\includegraphics[scale=0.39]{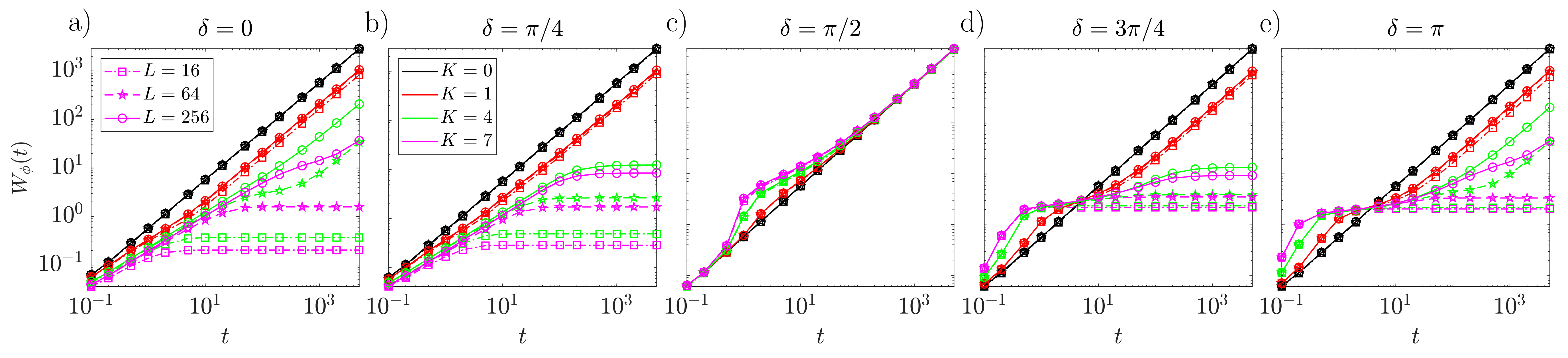}
\vspace{-0.7cm}
\caption{{\sf \bf Roughness $W_\phi(L,t)$ as a function of time for the oscillator lattice defined by Eq.\ (\ref{eqsim}) for different values of $\delta$ in each panel:} (a) $\delta = 0$, (b) $\delta = \pi/4$, (c) $\delta = \pi/2$, (d) $\delta = 3\pi/4$, and (e) $\delta = \pi$. For each $\delta$ we consider sizes $L=16$, 64, and 256 (denoted by different symbols and line styles) and coupling strengths $K = 0,1,4,7$ (denoted by different colors); see the legends in panels (a) and (b), which are valid for all panels. Results are averages over $200$ realizations (i.e., assignments of the intrinsic frequencies). }\label{figsat}
\end{figure*}

We focus on $\delta \in (-\pi/2,\pi/2)$ ($\cos \delta > 0$) in Fig.~\ref{figPD}, as it is only there that the small-slope approximation $\Delta \phi \ll 1$ can be expected to hold. The two oscillators synchronize for $K\geq K_c^\text{SSK}$, with a stationary $\Delta \phi$ (whose sign is given by that of $\Delta \omega$), that gets closer to $0$ with increasing $K$. They evolve at the common frequency
\begin{equation}
\omega_1^\text{eff} = \omega_2^\text{eff} = \bar{\omega} - K \sin (\delta) \left(1-\sqrt{\frac{K^2 - (K_c^\text{SSK})^2}{K^2}}\right),
\label{omegaeff2osc}
\end{equation}
which, for $\delta \neq 0$, is shifted away from the average of the intrinsic frequencies $\bar{\omega}$ by an amount that is proportional to $-\sin \delta$. While this two-oscillator model cannot provide an accurate picture of the many-oscillator system, it is in fact quite useful for discussing various points related to the stationary slope distribution and the general phenomenology of our model, as will be shown below.

In Ref.~\cite{ostborn} the full many-oscillator problem is analyzed, and synchronization is shown to be possible above a given coupling strength $K_c^\text{O} \equiv (\omega_\text{max}-\omega_\text{min})/|d(\widetilde{\Delta \phi})|$. Here $d(\Delta \phi) = [\Gamma(\Delta \phi) + \Gamma(-\Delta \phi)]/K =2\sin\delta (\cos{\widetilde{\Delta \phi}}-1),$ where in the last step we have particularized the coupling function to the one given in Eq.~(\ref{gamma}). The phase difference or slope $\widetilde{\Delta \phi}$ is defined to be the largest value such that $\Gamma(\Delta \phi)$ is increasing for all $\Delta \phi < |\widetilde{\Delta \phi}|$ [as our $\Gamma(\Delta \phi)$ always is in some neighborhood of the origin for $\delta \in (-\pi/2,\pi/2)$]. We find $\widetilde{\Delta \phi} = \pi/2-\delta$ for $\delta \in (0,\pi/2)$ and $-\pi/2-\delta$ for $\delta \in (-\pi/2,0),$ which yields $d(\widetilde{\Delta \phi}) = -2 \sin \delta (1-|\sin\delta|)$. Hence, the estimate of the critical coupling strength in our model,
\begin{equation}
K_c^\text{O} = \frac{1}{|\sin \delta| (1-|\sin\delta|)},
\label{ko}
\end{equation}
which is shown as a blue dotted line in Fig.~\ref{figPD}. In principle this estimate works for an arbitrarily large system.

Both $K_c^\text{SSK}$ and $K_c^\text{O}$ show a divergence of the critical coupling as $\delta \to \pm \pi/2$, see Fig.~\ref{figPD}, which will be important in the discussion of the following subsection. $K_c^\text{O}$ can be expected to work better, as it is based on a sophisticated analysis of the full many-oscillator problem. Furthermore, it diverges for $\delta \to 0$, as $K_c$ is known to  do in the thermodynamic limit \cite{strogatz}. On the other hand, $K_c^\text{O}$ appears to overestimate $K_c$ in some cases \cite{ostborn}, and possible discrepancies have been found in a recent contribution \cite{moroney}. The remaining information displayed in Fig.~\ref{figPD} will be discussed at the end of the following subsection.

\subsection{Saturation and stationary slopes: Parameter choices}

\label{satsta}

We next explore numerically the existence of synchronous motion in our many-oscillator model, and its dependence on $\delta$ and  the system size $L$. In Fig.~\ref{figsat} we show the roughness $W_\phi(L,t)$ as a function of time for the oscillator lattice defined by Eq.~(\ref{eqsim}), each panel containing results for one possible value of $\delta$, including $\delta = 0$, $\pi/4$, $\pi/2$, $3 \pi/4$, and $\pi$. In order to investigate the role played by the system size, we consider systems with $L=16$, $64$, and $256$ oscillators. Let us first focus the discussion on the case of odd coupling, $\delta = 0$ and $\pi$, panels (a) and (e), respectively. While synchronization, i.e.\ the saturation of $W_{\phi}(L,t)$, is indeed achieved for sufficiently large coupling strength $K$, the critical value clearly increases with the system size, as expected from the exact results for $\delta = 0$ \cite{strogatz}. Take for example the case of $K=4$ (green lines), which stabilizes the width for long times if $L=16$ (squares), but not for $L=64$ (stars) or $256$ (circles).

In contrast, for $\delta = \pi/4$ and $3\pi/4$, panels (b) and (d) in Fig.~\ref{figsat} respectively, the critical coupling strength seems to be independent of the system size, as argued in Ref.~\cite{ostborn}.
A qualitatively similar behavior is observed for $\delta = -\pi/4$ and $-3\pi/4$, as expected from the invariance mentioned at the end of Sec.\ \ref{nummod}, as well as for other inspected values in each of the four quadrants (not shown). As for $\delta = \pi/2$, i.e.\! the even-symmetry case $\Gamma({\Delta \phi}) - \Gamma({-\Delta \phi}) = 0$, considered in Fig.~\ref{figsat} (c), the behavior is completely different and synchronization is never achieved. In fact, the coupling between oscillators does not even reduce the rate of increase of the roughness with respect to the uncoupled ($K=0$) case, and the same behavior is observed for $\delta = -\pi/2$. Both estimates of the critical coupling strength $K_c$ considered in the previous subsection diverge for $\delta\to \pm\pi/2$, see Fig.~\ref{figPD}; in fact not even a system of two oscillators synchronizes, so this is expected.

These numerical results for the roughness $W_\phi(L,t)$ will be helpful in finding values of $K(>K_c)$ appropriate for the study of synchronous dynamics for different $\delta$ in the following. But before doing that, we need to determine the range of $\delta$ for which the small-slope approximation is expected to hold. According to the discussion in Sec.~\ref{contlim}, for such values of $\delta$ the dynamics is effectively described by the continuum approximation in Eq.~(\ref{eq4}), whose parameters ---which are related to those of the synchronization model (\ref{eqsim})  through the derivatives of the coupling function (\ref{gamma}) at $\Delta \phi = 0$--- are
\begin{equation}
\nu = a^2 K \cos \delta,\ \ \frac{\lambda}{2}= - a^2 K \sin \delta.
\label{gammaders}
\end{equation}
The range of validity of this approximation is studied in detail in Appendix~\ref{App0}, and leads to the conclusion that it can only be valid for $\delta \in (-\pi/2,\pi/2)$, i.e. for $\cos \delta > 0$, see also the first condition in Eq.~(\ref{twoosccond}). On the other hand, the fact that the results reported in Fig.~\ref{figsat} (and also Fig.~\ref{figslopes} in Appendix~\ref{App0}) are unaffected by a change in the sign of $\delta$ (not shown) can be traced back to the invariance of the oscillator system under the global sign flip $\phi_i \to - \phi_i$ and $\delta \to -\delta$, discussed in Sec.\ \ref{nummod}. For $\cos \delta > 0$, when the continuum description is justified, this fact is possibly related with the known independence (at least for the case of time-dependent noise) of the large-scale behavior with respect to the sign of $\lambda$ \cite{barabasi,krug97}. Indeed, the only feature of the KPZ universality class that depends on this sign is that of the skewness of the (TW) fluctuation distribution \cite{kriecherbauer10,takeuchi}. Most of our results will explore the interval $[0, \pi/2)$, because this invariance under simultaneous phase and $\delta$ reversal makes $(-\pi/2,0)$ redundant.


Thus, in the following, we focus our numerical analysis on values of $K(>K_c)$ large enough so that synchronization is achieved with sufficiently small stationary slopes $\Delta \phi$, for $\delta$ taking values through the interval $[0,\pi/2)$. This is in principle always possible, though for $\delta = 0$ it requires choosing some $K$ that grows with the system size as $\sqrt{L}$ \cite{strogatz}. To that end, we will consider coupling strengths $K$ that are roughly twice as large as our numerical estimate for $K_c$. The latter is defined as the smallest value of $K$ for which we obtain a saturation of the roughness $W_\phi(L,t)$ for $200$ realizations (as many as those employed in the results reported in Fig.~\ref{figsat}); see Appendix \ref{AppDep} for more details. Such numerical estimates, which are likely to be lower bounds for the actual value of $K_c$, are shown in Fig.~\ref{figPD} for different system sizes $L$ (see legend). They are consistently larger than the two-oscillator estimate $K_c^\text{SSK}$ (\ref{kssk}), as expected, and smaller than $K_c^\text{O}$ (\ref{ko}), in agreement with previous results \cite{ostborn}. For the case of $\delta = 0$, the estimate does grow with $\sqrt{L}$, while in the other cases it appears to be only weakly dependent on $L$. A complementary analysis of the dependence of $K_c$ on $L$ for different $\delta$, including some additional results, is provided in Appendix \ref{AppDep}.

\section{Synchronization as a growth process for Kuramoto coupling ($\delta = 0$)}
\label{odd}


In this section we address the numerical solution of Eq.~(\ref{eqsim}) for $\delta = 0$, which becomes
\begin{equation}
\frac{d\phi_i}{dt} = \omega_i +  K [\sin(\phi_{i+1}\!-\!\phi_i) +\sin(\phi_{i-1}\!-\!\phi_i)].
\label{eqsimodd}
\end{equation}
This is the Kuramoto model in a one-dimensional lattice, which is known not to synchronize in the thermodynamic limit, $L\to\infty$ \cite{strogatz}. Since we are nonetheless interested in systems of finite size, we will focus on a value of $K$ that is large enough to achieve synchronization for the size under consideration. By studying the saturation of $W_\phi(t)$, we find $K_c \gtrsim 0.6 \sqrt{L}$ based on systems with sizes ranging from $L = 16$ to $512$ (see Appendix \ref{AppDep}). The transition to synchronization in higher-dimensional lattices of Kuramoto oscillators has been studied in Refs.\! \cite{hong,hong2007}.

\begin{figure}[t!]
\includegraphics[scale=0.47]{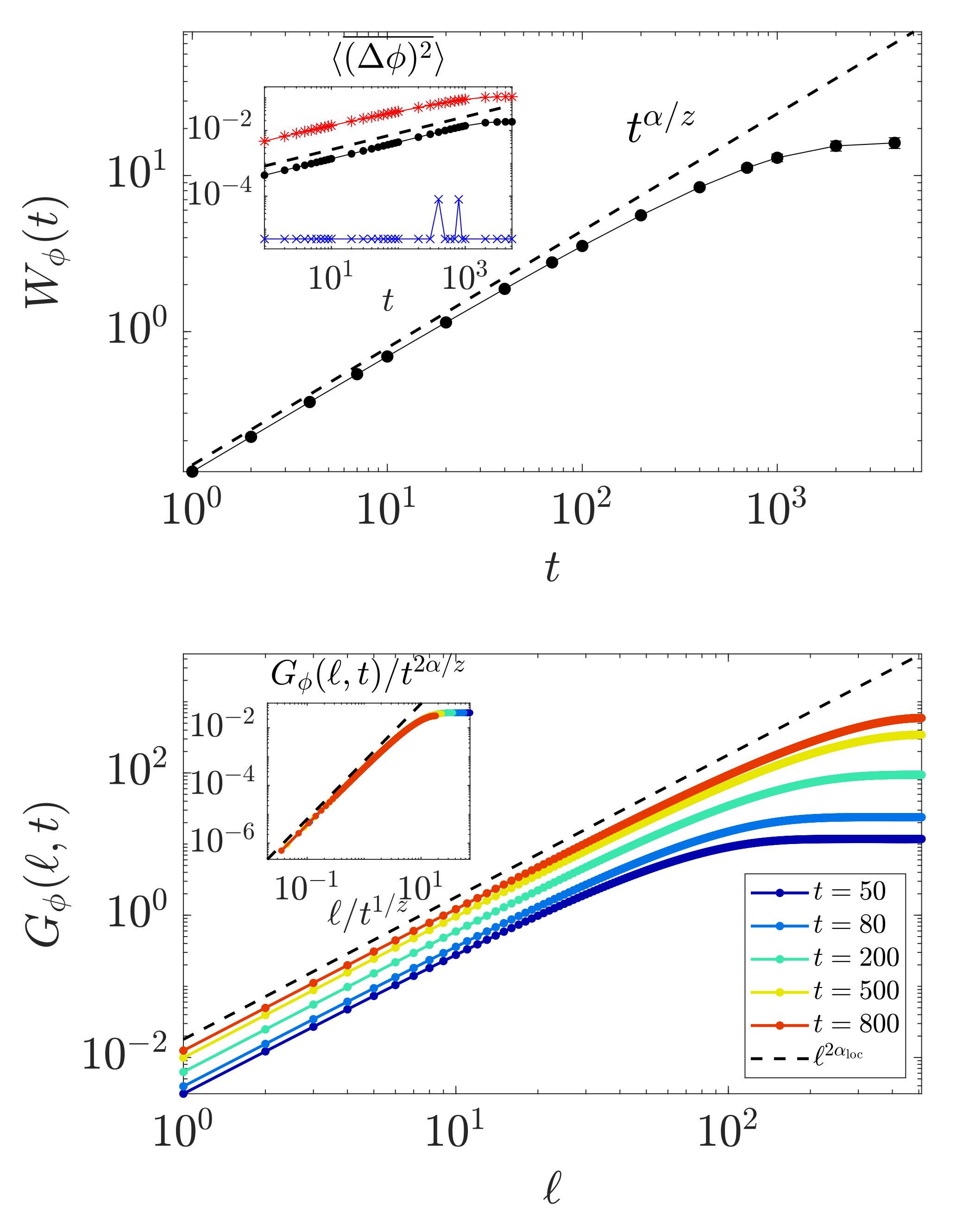}
\vspace{-0.3cm}
\caption{{\sf \bf Roughness $W_\phi(t)$ and height-difference correlation function $G_\phi(\ell,t)$ for the Kuramoto model, Eq.\  (\ref{eqsimodd}) ($\delta = 0$) with size $L = 1024$ and coupling strength $K= 40$.}
({\it Upper panel}) $W_\phi(t)$ as a function of time in the synchronous regime, with error bars displaying the standard error of the average across realizations. The inset shows the average squared slope $\langle\overline{(\Delta \phi)^2} \rangle$ (black circles), as well as the average of the maximum squared local slope (red asterisks), and the most probable value of the squared slope (blue crosses), as a function of time for the same parameter values. The black dashed line shows the power-law growth of the average squared slopes predicted by the super-rough scaling Ansatz, Eq.\ \eqref{glt_mod}, as $t^{2(\alpha-\alpha_\text{loc})/z}$. ({\it Lower panel}) $G(\ell,t)$ as a function of $\ell$ in the synchronous regime for different times (see legend). The dashed line shows the power law growth with the distance $\ell^{2\alpha_\text{loc}}$. Inset: Rescaling of $G(\ell,t)$ following the theoretical form in Eq.~(\ref{glt_mod}). In all power-law visual guides we have used the Larkin model values $\alpha = 3/2$ and $z=2$, as derived from Eq.\ \eqref{SkLarkin}, and $\alpha_\text{loc}=1$ due to the super-rough scaling. Averages based on 2000 realizations.
}\label{figWGdelta0K40}
\end{figure}

\subsection{Kinetic roughening: Scaling Ansatz and exponents}

In the upper panel of Fig.~\ref{figWGdelta0K40} we show the roughness $W_\phi(t)$ of a system of $L= 1024$ oscillators with coupling strength $K=40$. According to the numerical estimate mentioned above, the critical coupling strength in this case is $K_c\gtrsim  20$, so the strength of choice is roughly twice as large. A clear region of growth with exponent $\beta = \alpha/z = 3/4$, is observed, followed by a saturation that consolidates around $t\approx 10^3$. As indicated in Table \ref{tab}, this is consistent with the exponent values of the Larkin model, $\alpha = 3/2$ and $z = 2$.

In the inset we show that the average squared slope $\langle\overline{(\Delta \phi)^2} \rangle$ (black circles) increases as  $t^{2(\alpha-\alpha_\text{loc})/z} = t^{1/2}$ (where again the Larkin model exponent $\alpha_\text{loc} = 1$ ensues; see the black dashed line) until the growth saturates \cite{lopez99}. Indeed, this is one of the hallmarks of anomalous scaling \cite{dassarma,schroeder93,lopez}: the local slopes, instead of stabilizing rapidly as the correlation length $\xi(t)$ exceeds their length scales, keep on increasing with time until saturation, $\xi(t)\sim L$. This increase is given by  Eq.\ (\ref{glt_mod}) when the spectral roughness exponent $\alpha_s > 1$, which is shown to be the case below. Such a saturation of the slopes may provide a natural
explanation for the fact that one-dimensional systems of oscillators with Kuramoto coupling $\Gamma(\Delta \phi) = K \sin(\Delta \phi)$ do not synchronize in the thermodynamic limit, where saturation can never be achieved \cite{strogatz}. The average of the maximum squared local slope, i.e.\! the average across realizations of the maximum squared slope in the system, see also the inset of the upper panel of Fig.~\ref{figWGdelta0K40} (red asterisks), also increases similarly. We also report the most probable value of the squared slope (blue crosses), which always stays close to zero, as corresponds to an asymptotic unimodal distribution with zero mean, see Fig.~\ref{figslopes}(a) below. This feature, which is also explored in Appendix \ref{AppA}, will reveal a sharp distinction between the Kuramoto coupling ($\delta = 0$) and other couplings for $\delta \neq 0$ in the next section and in the appendices.

A more detailed view on the two-point correlations beyond nearest neighbors can be obtained by inspection of the full height-difference correlation function $G_\phi(\ell,t)$, shown in the lower panel of Fig.~\ref{figWGdelta0K40}. The correlation function behaves as expected from a system displaying super-rough anomalous scaling, see Eq.~(\ref{glt_mod}), with a dependence on distance of the form $\ell^2$ that reveals a local exponent $\alpha_{\text{loc}} = 1$, and a growth in time that proceeds until the width reaches the saturation regime. In the inset we show $G_\phi(\ell,t)/t^{2 \alpha/z}$ as a function of a rescaled time variable $u= \ell/t^{1/z}$, which scales as $u^{2 \alpha_\text{loc}}$ for $u\ll 1$ and reaches a constant value for $u\gg 1$, as expected from Eq.\ \eqref{glt_mod}.

This is all in agreement with the analysis of the continuum description. Indeed, from the discussion in Sec.\ \ref{contlim}, the small-slope approximation for odd coupling functions yields for the Kuramoto model
\begin{equation}
\partial_t \phi({\bf x},t) =   \omega({\bf x}) + \nu \nabla^2 \phi({\bf x},t),
\label{eq6}
\end{equation}
where the surface tension $\nu$ is defined as in Eq.~(\ref{gammaders}). Notably, now the KPZ nonlinearity is absent, as $\lambda \propto \Gamma^{(2)}(0) = 0$. Equation (\ref{eq6}) is the Larkin model, namely, the EW equation with columnar noise, in which $\nu> 0$ plays the role of an elastic constant. Some of its main features from a kinetic roughening perspective are reviewed in Ref.~\cite{purrello}. This equation was first proposed in the study of pinning of vortex lines in type II superconductors \cite{larkin}. In the synchronization context it was considered in Ref.~\cite{hong}.

\begin{figure}[h!]
\includegraphics[scale=0.47]{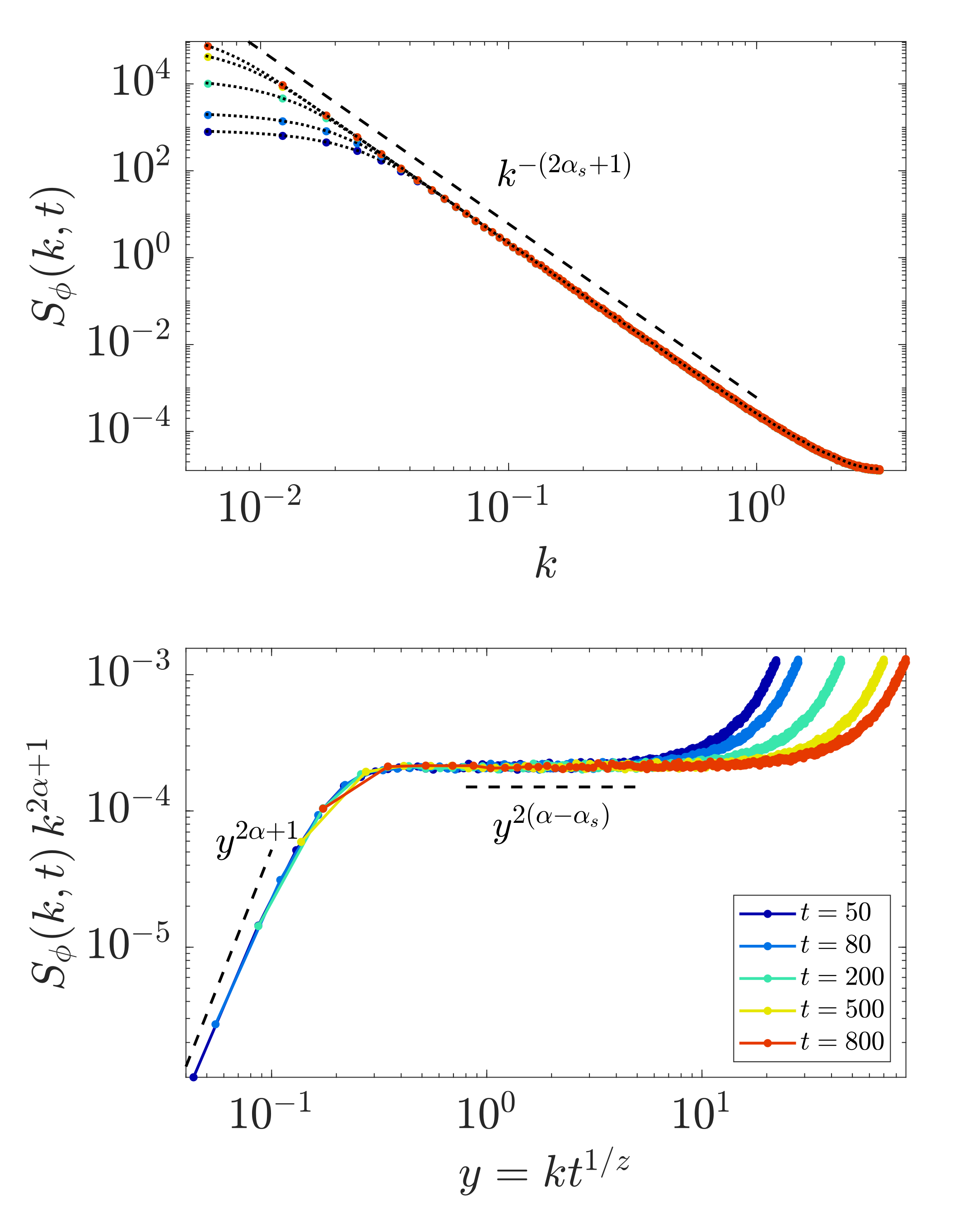}
\vspace{-0.3cm}
\caption{{\sf \bf Structure factor $S_\phi(k,t)$ as a function of time for the Kuramoto model, Eq.\ (\ref{eqsimodd}), with $L = 1024$, and $K= 40$.} ({\it Upper panel}) $S_\phi(k,t)$ as a function of $k$ (colored dots) and analytical expression for the Larkin model, Eq.\ \eqref{SkLarkin} (dotted lines). ({\it Lower panel}) $k ^{2\alpha+1} S_\phi(k,t)$ as a function of $y = k t^{1/z}$. In all rescalings and power-law visual guides, $\alpha = \alpha_s = 3/2$, and $z=2$. The structure factor curves are shown for the same time points displayed in the lower panel of Fig.~\ref{figWGdelta0K40}. Averages based on 2000 realizations.}
\label{figSkdelta0K40}
\end{figure}

The solution of Eq.\ \eqref{eq6} is straightforward in Fourier space, as the equation is linear.
For the initial condition of a homogeneous phase profile, ${\phi}({\bf x},0) = 0$, the exact structure factor reads \cite{purrello}
\begin{equation}
S_\phi({\bf k},t) = \langle \hat{\phi}({\bf k},t) \hat{\phi}(-{\bf k},t)\rangle =  \frac{(2\pi)^d 2 \sigma}{\nu^2 k^4} \left(1-e^{-\nu k^2 t} \right)^2.
\label{SkLarkin}
\end{equation}
Comparing with the generic scaling ansatz in Eq.~(\ref{Sk}), the roughness exponents are $\alpha = \alpha_s = 3/2$ and the dynamic exponent is $z=2$ \cite{ramasco,purrello}. Given that this structure factor follows the FV scaling form, Eq.\ (\ref{SkFV}), it corresponds to a system with super-rough anomalous scaling in which $\alpha_{\rm loc}=1$.
Actually, the analytical solution of the Larkin model, Eq.\ \eqref{SkLarkin}, describes very accurately the structure factor calculated in simulations of the Kuramoto model, Eq.\ \eqref{eqsimodd}. Results of the latter are provided in Fig.\ \ref{figSkdelta0K40}, which shows the numerical $S_\phi({\bf k},t)$ as a function of $k$ for different values of time.

In the upper panel, the structure factor of the Larkin model, Eq.~(\ref{SkLarkin}) (represented as black dotted lines), shows excellent agreement with the simulation data of the oscillator lattice. We have particularized the analytical expression for $\nu = \Gamma^{(1)}(0) = K$, the overall constant factor, which includes the (unknown) noise strength $\sigma$ in the continuum approximation, being manually adjusted to the data but time-independent. The agreement for large $k$ values improves by replacing the wave number $k$ by $2 \sin(k/2)$, which makes the structure factor a periodic function of $k$, as implied by the existence of a finite wave-vector cut-off in the oscillator lattice \cite{siegert96}. This replacement has very little effect at small $k$. Once again, visual guides (represented as black dashed lines) based on the scaling form, Eq.\ (\ref{Sk}) with the theoretical exponents corresponding to FV scaling with super-rough behavior are shown to work extremely well, including the spectral roughness exponent $\alpha_s = 3/2$.
This is further confirmed in the lower panel, where we perform a data collapse of the
$S(k,t)$ data which agrees with Eq.\ \eqref{Sk},
except for wave numbers in close proximity to the cutoff. The behavior just described for the structure factor is known to imply \cite{lopez} the anomalous scaling and exponent values of the roughness and height-difference correlation functions assessed in Fig.\ \ref{figWGdelta0K40}.



\subsection{Dynamics of synchronization}

\begin{figure}[t!]
\includegraphics[scale=0.36]{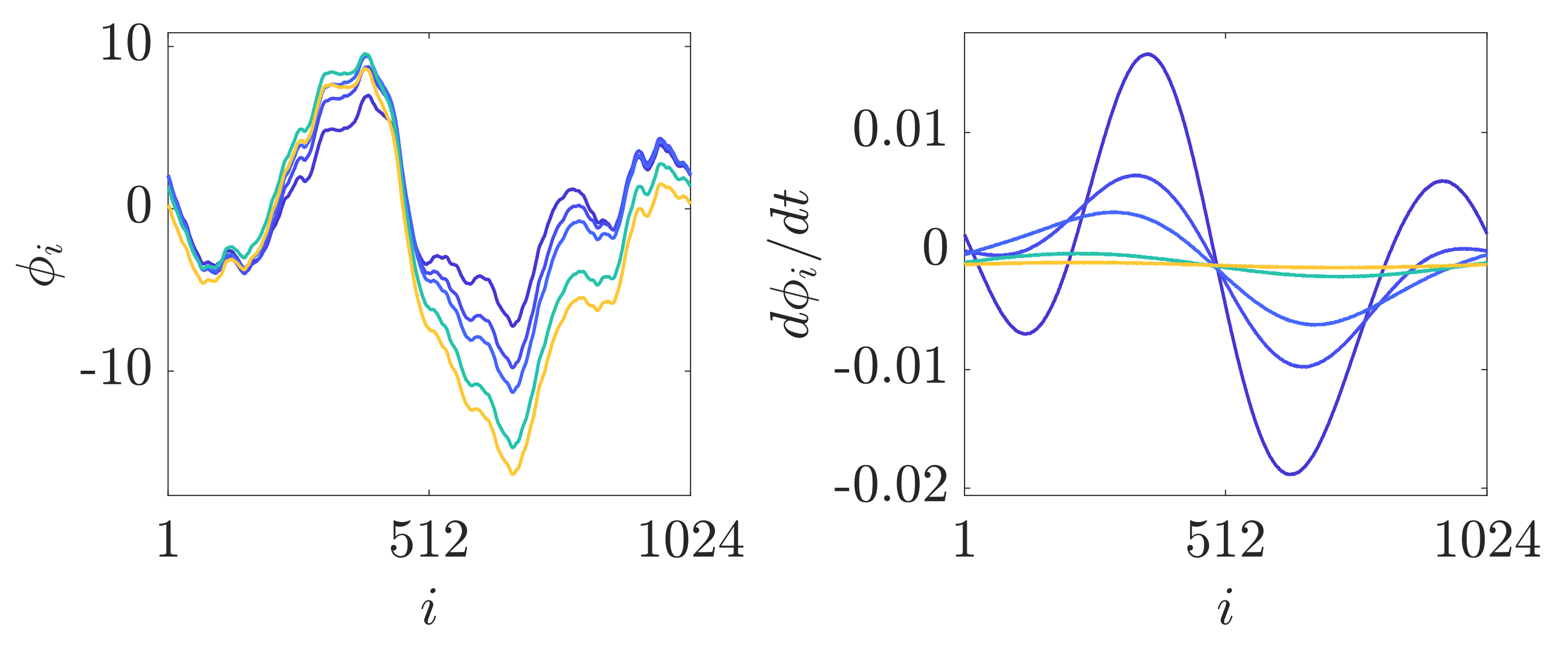}
\vspace{-0.7cm}
\caption{{\sf \bf Phases $\phi_i$ and instantaneous frequencies $d\phi_i/d t$ across time in a representative trajectory of the Kuramoto model, Eq.\ (\ref{eqsimodd}), with $L = 1024$ and $K= 40$.} ({\it Left panel}) Phases $\phi_i$ for $t = 200, 400, 600, 1600$, and $2600$. Earlier times are shown in darker colors, later times in lighter colors. ({\it Right panel}) Instantaneous frequencies $d{\phi}_i/dt$ for the same trajectory, at the same time points.}
\label{FigTrajdelta0K40}
\end{figure}

To shed light on the phenomenology underlying this super-rough scaling, in Fig.~\ref{FigTrajdelta0K40} we show a representative trajectory (see file S1 within the Supplemental Material \cite{suppl} for a movie). Specifically, we show the phases $\phi_i$ (left panel) and instantaneous frequencies $d\phi_i/d t$  (right panel) as functions of time, with darker colors indicating earlier times and lighter colors later times. At the initial time considered ($t=200$) we find a configuration where important differences in the instantaneous frequencies exist. These differences lead to oscillators evolving at different rates, so that the `interface' becomes rougher. As a result of this roughening, however, there is a progressive reduction of the differences in the instantaneous frequencies, which eventually leads to a synchronous dynamics.

The evolution appears to be quite modest and slow, in the sense that, apart from slight rearrangements of the slopes, the main motion is a uniform displacement at the average frequency $\omega^\text{eff}$ that is eventually attained by all the oscillators. Indeed, if one performs the space average of Eq.~(\ref{eq6}), due to the periodic boundary conditions and to the fact that the mean intrinsinc frequency $\langle \omega \rangle$ is zero, one obtains a vanishing interface velocity. For time-dependent noise, this is a standard feature of EW surface growth \cite{barabasi}. At the discrete level, from Eq.~(\ref{eqsimodd}) we obtain, after a slight manipulation of the indices,
\begin{equation}
\overline{\frac{d\phi_i}{d t}} = \overline{\omega_i }  +  K \overline{[\sin(\phi_{i+1}\!-\!\phi_i)+\sin(\phi_i\!-\!\phi_{i+1})]}= \overline{\omega_i}.
\label{meanfreqodd}
\end{equation}
The odd symmetry of the function $\Gamma(\Delta \phi) = \sin(\Delta \phi)$ makes the coupling term vanish in Eq.\  (\ref{meanfreqodd}). If the system is sufficiently large, the average instantaneous frequency becomes zero, $\overline{\frac{d \phi_i}{dt} } = \overline{\omega_i} \approx \langle \omega \rangle = 0$, by the law of large numbers.

\subsection{Fluctuation statistics}
\label{sec:flKuramoto}

\begin{figure}[h!]
\includegraphics[scale=0.50]{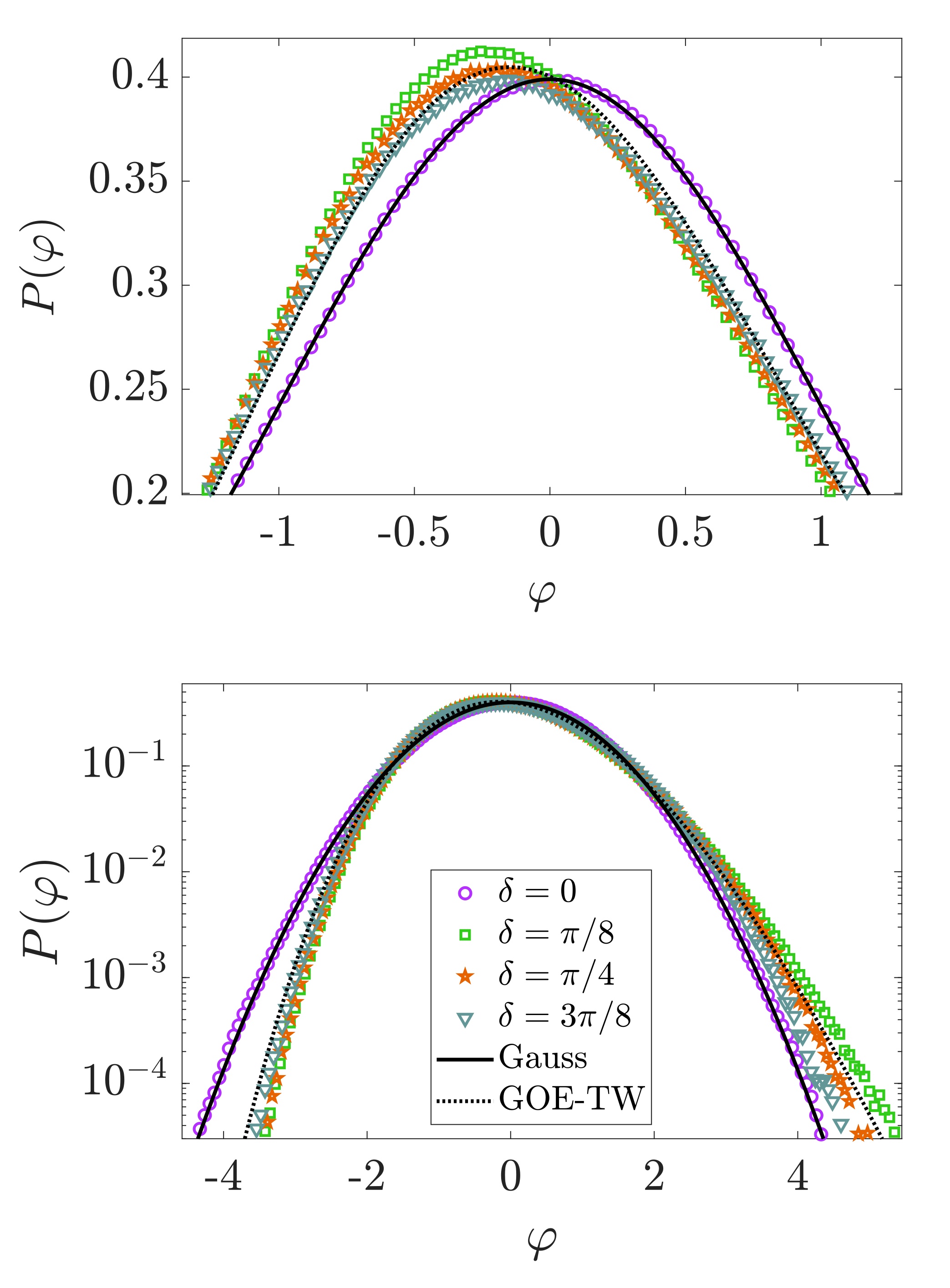}
\vspace{-0.3cm}
\caption{{\sf \bf Histogram of fluctuations $\varphi_i$ defined as in Eq.\ \eqref{fluct} for the oscillator lattice, Eq.\ \eqref{eqsim}, in a system of $L=8192$ oscillators, for $\delta = 0$ ($K=110$), $\delta = \pi/8$ ($K=4$), $\delta = \pi/4$ ($K=4$), and $\delta = 3 \pi/8$ ($K=8$).} ({\it Upper panel}) Linear scale. ({\it Lower panel}) Logarithmic scale. We always take as reference time $t_0 = 50$, and values of $\Delta t = 50, 100, 150, \ldots,500$, which are in the interval where a power-law growth $W_\phi(t) \sim t^\beta$ is observed. The black solid line corresponds to a Gaussian distribution, while the dotted line shows a GOE-TW distribution, see Sec.\ \ref{nonodd}. Histograms and theoretical curves have been normalized to zero mean and unit variance. No fitting parameters are used. Histograms based on $10^4$ realizations.}
\label{FigDistalldeltas}
\end{figure}

Finally, beyond scaling exponents and the general phenomenology of trajectories, we also explore the statistics of the phase fluctuations, as recent developments in kinetic roughening show that the fluctuation PDF and covariance are also essential traits to define a universality class, see e.g.\ Refs.\ \cite{kriecherbauer10,takeuchi,rodriguez-fernandez21,marcos22} and others therein. In Fig.~\ref{FigDistalldeltas} we show the numerically-obtained PDF of the fluctuations $\varphi_i$ defined as in Eq.~(\ref{fluct}) for several values of $\delta$. Our discussion for the moment is focused on the case $\delta = 0$, see the magenta circles. A larger size, namely $L=8192$ (with $K = 110$, which again is roughly twice as large as the estimated $K_c$) is considered for these observables, which are especially affected by finite-size effects. An initial time of $t_0 = 50$ and time windows $\Delta t = 50, 100, \ldots, 500$ have been chosen, which correspond to the intermediate growth regime. In Fig.~\ref{FigDistalldeltas} the same distribution is shown in linear (upper panel) and in logarithmic scales (lower panel). The histogram has been rescaled and shifted so that the sample mean is zero and the sample standard deviation is one. The purpose is to make the comparison with a normalized Gaussian PDF, which is shown with a black solid line. A very good agreement is found between the numerical results and the theoretical distribution without any kind of fit, which confirms expectations based on the good agreement between the Kuramoto and the Larkin models, and the fact that the latter is linear. Indeed, for time-dependent noise the EW equation is another well-known instance of Gaussian height fluctuations \cite{barabasi,krug97}.

\begin{figure}[h!]
\includegraphics[scale=0.40]{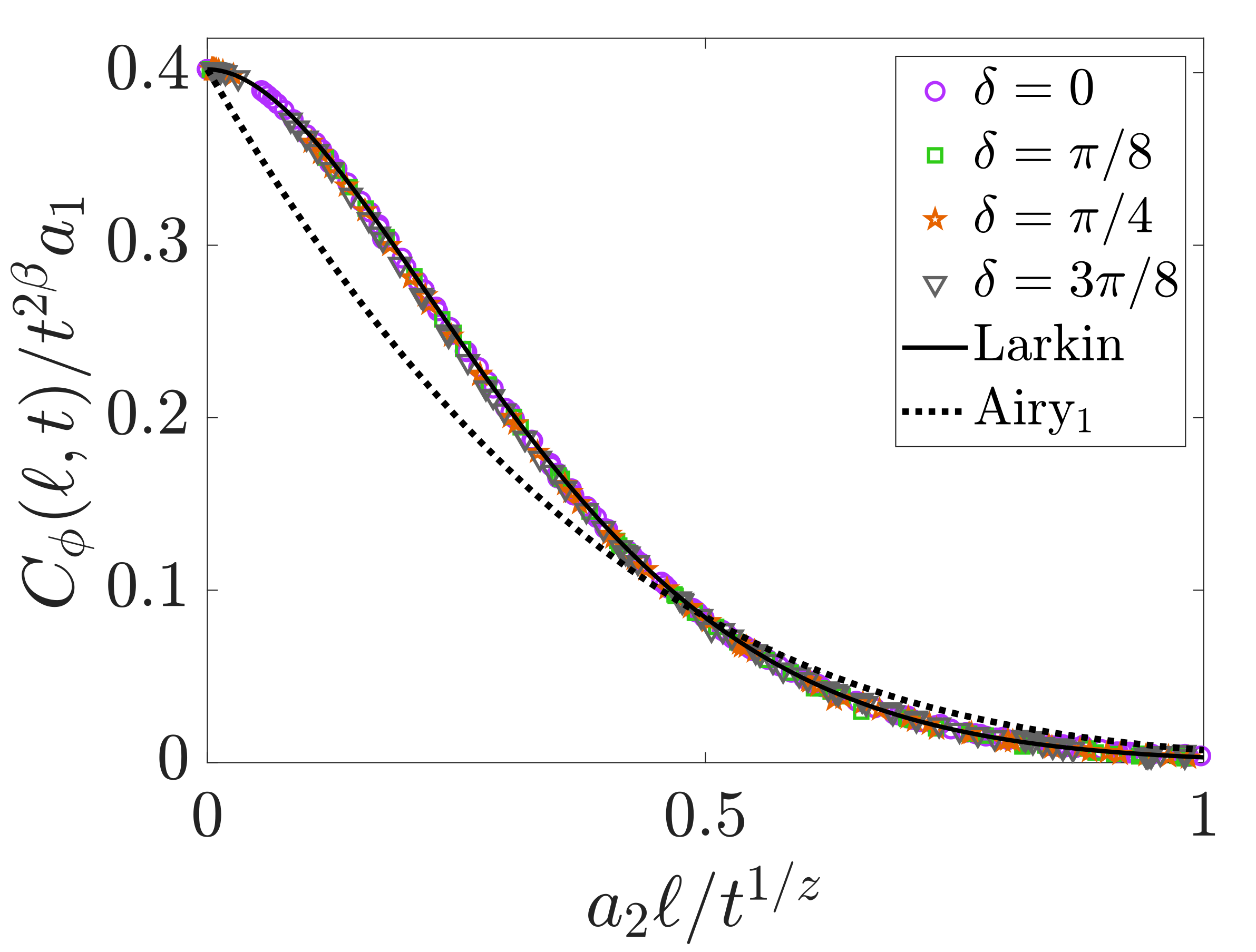}
\vspace{-0.4cm}
\caption{{\sf \bf Rescaled covariance $C_\phi(\ell,t)$ for the oscillator lattice, Eq.\ \eqref{eqsim}, in a system of $L=8192$  oscillators for $\delta = 0$ ($K=110$), $\delta = \pi/8$ ($K=4$), $\delta = \pi/4$ ($K=4$), and $\delta = 3 \pi/8$ ($K=8$).} The time points considered are in the growth regimes and coincide with those included in Fig.~\ref{FigDistalldeltas}. The axes have been rescaled according to Eq.~(\ref{covscal}). The black solid line corresponds to the exact covariance of the Larkin model, Eq.\ \eqref{eq:cLark}, while the dotted line shows the covariance of the Airy$_1$ process which has been used in the determination of the rescaling factors $a_1$ and $a_2$, see the text. Numerical covariances based on $10^3$ realizations.}
\label{FigCovariances}
\end{figure}

One further ingredient of the fluctuation statistics is the phase covariance $C_\phi(\ell,t)$. Results for this correlation function based on the numerical solution of Eq.\ \eqref{eqsim} are shown in Fig.~\ref{FigCovariances} for the same values of $\delta$ and the same time points employed in Fig.~\ref{FigDistalldeltas}. We again focus on $\delta = 0$ (Kuramoto model) for the time being, see the magenta circles. In the growth regime, the covariance should scale as \cite{kriecherbauer10,takeuchi}
\begin{equation}
C_\phi(\ell,t) = a_1 t^{2\beta} \mathcal{C}(a_2 \ell/t^{1/z}),
\label{covscal}
\end{equation}
where $\mathcal{C}$ is a scaling function. The constants $a_1$ and $a_2$ are determined with respect to a reference value $x_0$ based on an assumed functional form for $\mathcal{C}(x)$, so that all curves pass through $(x_0, \mathcal{C}(x_0))$, see Ref.~\cite{barreales2020} for further details.
We have chosen the covariance of the Airy$_1$ process \cite{takeuchi} as the functional form for $\mathcal{C}(x)$, which is shown as a black dotted line, a choice whose significance will be discussed in the next section.

For the Kuramoto model ($\delta = 0$), the covariance again follows accurately the analytical curve of the Larkin model, which differs substantially from Airy$_1$. Indeed, since $C_\phi(\ell,t)$ is the inverse Fourier transform of the structure factor \cite{siegert96}, using Eq.\ \eqref{SkLarkin} the covariance of the Larkin model can readily be shown to comply with a scaling form like that described by Eq.\ \eqref{covscal} with $\beta=3/4$ and $z=2$, but with a scaling function which differs from Airy$_1$, namely,
\begin{equation}
    C_\phi(\ell,t) = \frac{4 \pi \sigma t^{3/2}}{\nu^{1/2}} \mathcal{F}^{-1}\!\left[\frac{\left(1-e^{-\kappa^2} \right)^2}{\kappa^4}\right](\ell/\sqrt{\nu t}) ,
    \label{eq:cLark}
\end{equation}
where $\kappa=\sqrt{\nu t} k$.
This analytical form for the covariance of the Larkin model is shown in Fig.~\ref{FigCovariances} as a black solid line, and shows an excellent agreement with the numerical data for the oscillator lattice with $\delta = 0$ across the growth interval. Surprisingly, it also shows an excellent agreement with the covariances obtained for other values of $\delta$ (which are also displayed in Fig.~\ref{FigCovariances}) across their corresponding growth regimes, despite the crucial differences found for all other observables. This point will be further discussed in the next section.

\section{Synchronization as a growth process for non-odd coupling}
\label{nonodd}


In the continuum approximation  of the general oscillator model discussed in Sec.\ \ref{contlim}, if the coupling function is not odd,  $\Gamma(\Delta \phi)+ \Gamma(-\Delta \phi) \neq 0$, in general no derivatives of $\Gamma(\Delta \phi)$ can be assumed to vanish, and the dominant contributions in Eq.~(\ref{eq3}) are those of the KPZ equation with columnar noise, Eq.~(\ref{eq4}). This equation can be formally related to a diffusion problem in the presence of random traps and sources \cite{halpinhealy}. It is especially in that form that the equation has been studied, showing sharply localized solutions around pinning centers, which eventually hop to the basin of a more attracting center. This hopping dynamics proceeds across time scales, with longer times being spent at each new localization center that is attained. The connection between such form of localization and synchronization was first discussed a long time ago \cite{sakaguchi}, though it appears to be only very recently that the subject has been more thoroughly elucidated \cite{moroney}.

For the study of anomalous scaling in synchronous oscillator lattices, the main interest lies in the KPZ equation with columnar noise itself, which 
has been previously studied in the kinetic roughening literature \cite{szendro}. The localization of the diffusion problem is there shown to translate into the formation of facets, i.e.\! regions of nearly constant slope. Specifically, there is a coarsening process whereby some facets, which move uniformly at constant velocity, absorb other facets. In a finite-size system, the asymptotic solution is a triangular-looking surface growing at constant velocity. As will be seen next, in the oscillator problem the facets correspond to clusters of synchronized oscillators, and the coarsening process leads to larger (and faster) synchronized clusters, as previously observed in some models of oscillators in one and two dimensions \cite{sakaguchi,moroney}.

\subsection{Kinetic roughening: Scaling Ansatz and exponents}
\label{sec:exp_nonodd}

As we did in Sec.\ \ref{odd}, we study numerically a system of $L$ oscillators governed by Eqs.~(\ref{eqsim}), but this time $\delta = \pi/4$ (instead of $\delta = 0$). Other choices of $\delta\in (0,\pi/2)$, namely $\delta = \pi/8$ and $3\pi/8$, are considered in Appendix \ref{AppB} and lead to similar conclusions. The system synchronizes for $K$ above a critical value that is roughly $K_c\gtrsim  2$, according to the procedure described at the end of Sec.\ \ref{satsta}, see also Appendix \ref{AppDep}. Therefore, we will take as the coupling strength $K = 4$, which is approximately twice as large as the critical value.

\begin{figure}[h!]
\includegraphics[scale=0.47]{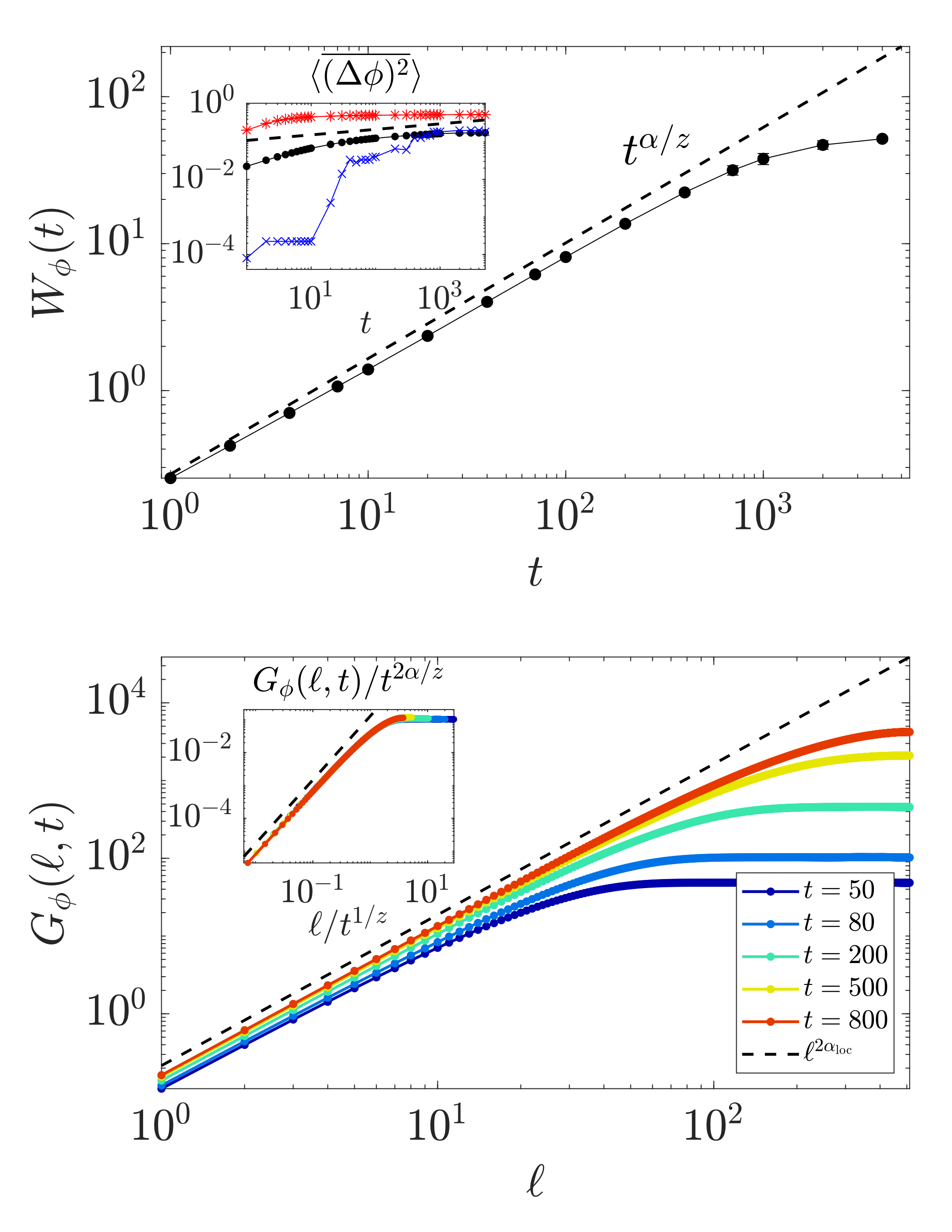}
\vspace{-0.3cm}
\caption{{\sf \bf Roughness $W_\phi(t)$ and height-difference correlation function $G_\phi(\ell,t)$ for the oscillator model defined by Eq.\ (\ref{eqsim}), with $\delta = \pi/4$ (non-odd coupling), $L = 1024$, and $K= 4$.}
({\it Upper panel}) $W_\phi(t)$ as a function of time in the synchronous regime, with error bars displaying the standard error of the average across realizations. In the power-law visual guides we have used the exponent values $\alpha = 1.07$ and $z=1.36$. The inset shows the average squared slope $\langle\overline{(\Delta \phi)^2} \rangle$ (black circles), as well as the average of the maximum squared local slope (red asterisks), and the most probable value of the local slope (blue crosses), as functions of time for the same parameter values. The black dashed line shows the power-law growth of the average squared slopes predicted by Eq.\ \eqref{glt_mod}, $t^{2(\alpha-\alpha_\text{loc})/z}$, for $\alpha_\text{loc}=0.97$. ({\it Lower panel}) $G(\ell,t)$  as a function of $\ell$ in the synchronous regime for different times (see legend). The dashed line shows the power-law growth with distance $\ell^{2\alpha_\text{loc}}$. Inset: Data collapse of $G(\ell,t)$ following the full theoretical form of Eq.~(\ref{glt_mod}). Averages based on 2000 realizations.
}\label{figWGdelta0250K4}
\end{figure}

In what follows, we find that a consistent description of the scaling behavior obtained in our simulations can be achieved for faceted exponent values $\alpha = 1.07\pm0.05$, $\alpha_s = 1.40\pm0.05$, $z = 1.36\pm0.05$, and $\alpha_\text{loc} = 0.97\pm0.05$; these must be adjusted quite precisely in order to satisfy the various scaling constraints pertaining to the different observables, so indeed the quoted uncertainties are conservative.
Note also that the local roughness exponent $\alpha_\text{loc} = 1$ for faceted scaling \cite{ramasco} is an upper bound for the local exponent for geometrical reasons \cite{leschhorn}, so that the possibility of finite size effects distorting some of the exponents (for $\alpha_\text{loc}$, necessarily towards smaller values) should also be considered. Their possibly non-universal nature may also explain why these exponents differ somewhat from those of Ref.~\cite{szendro}, and also (though less so) from those found for other values of $\delta$ in Appendix \ref{AppB}.

In Fig.~\ref{figWGdelta0250K4} we show the roughness $W_\phi(t)$  and the height-difference correlation $G_\phi(\ell,t)$ for a system of $L=1024$ oscillators, from numerical simulations of Eq.\ \eqref{eqsim} with $\delta = \pi/4$ and $K=4$. All the elements, symbols, and colors have the same meaning as in Fig.~\ref{figWGdelta0K40}. In the upper panel, the roughness saturates for times on the order to $10^3$. In the inset we show that the average squared slope $\langle\overline{(\Delta \phi)^2} \rangle$ (black circles) hardly follows the power-law growth, $t^{2(\alpha-\alpha_\text{loc})/z}$ expected from Eq.\ (\ref{glt_mod}), and stabilizes at a much shorter time scale than that needed for the saturation of the roughness \cite{lopez99}. This somewhat puzzling behavior is anticipated by the early saturation of the average of the maximum squared local slope (inset, red asterisks). Indeed the latter quantity is a proxy for the evolution of the facet with the steepest slope, as will be illustrated below. The most probable value of the squared slope (inset, blue crosses) also behaves in a completely different way than it did for $\delta = 0$ (see the inset of the upper panel in Fig.~\ref{figWGdelta0K40}): at some point it leaves the very small values from which it started and rises close to the value of $\langle\overline{(\Delta \phi)^2} \rangle$ itself. This issue will be addressed further in Appendix \ref{AppA}.

A more detailed view on space correlations beyond nearest neighbors is provided by the full height-difference correlation function $G_\phi(\ell,t)$, shown in the lower panel of Fig.~\ref{figWGdelta0250K4}. A dependence on distance of the form $\ell^{2\alpha_\text{loc}}$ is observed, as well as a growth in time that proceeds until $W_\phi(t)$ reaches saturation, but it appears to slow down earlier than for $\delta = 0$. Full agreement with Eq.\ (\ref{glt_mod}) can be appreciated in the data collapse of $G_\phi(\ell,t)$, which is displayed in the inset.

\begin{figure}[h!]
\includegraphics[scale=0.47]{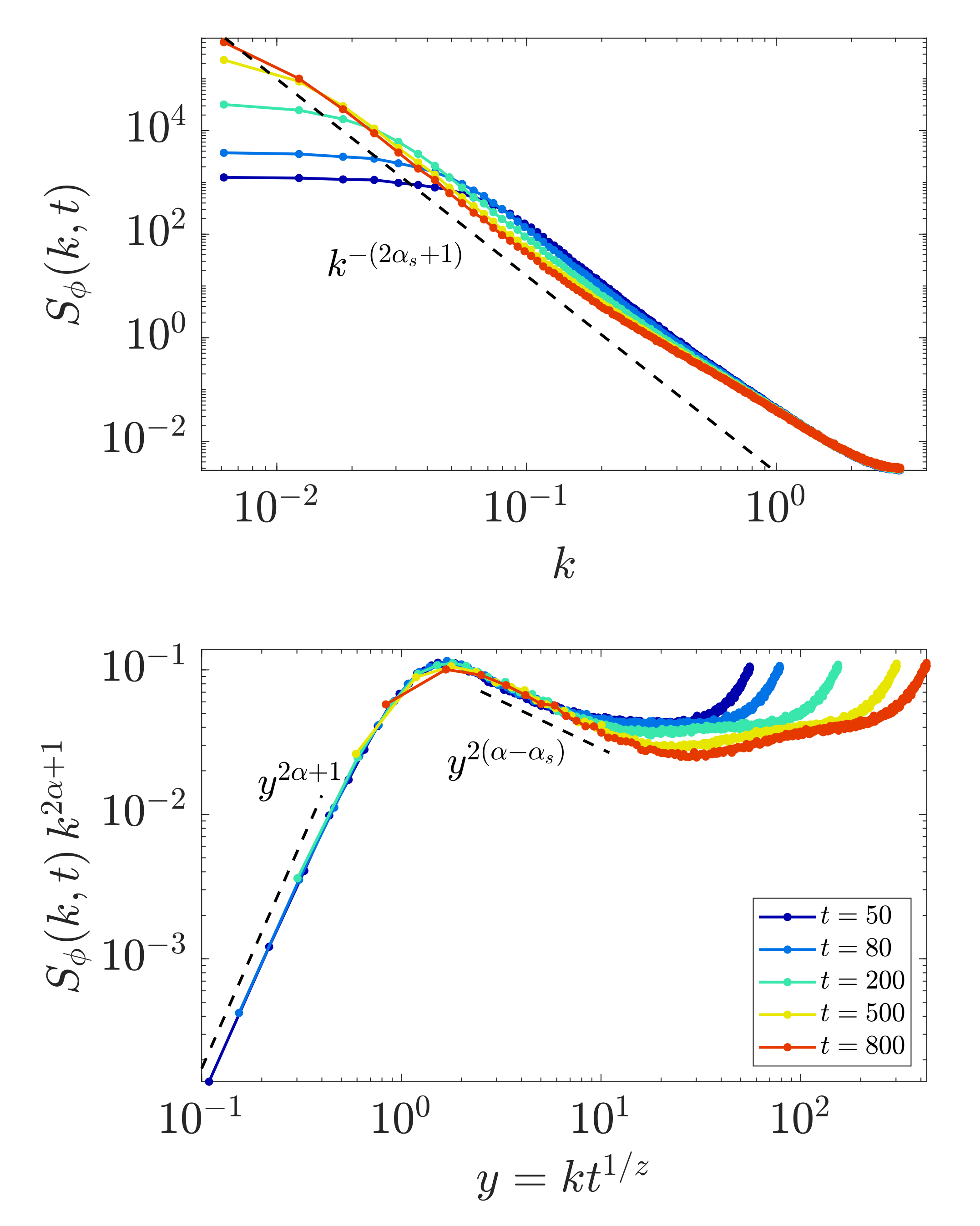}
\vspace{-0.3cm}
\caption{{\sf \bf Structure factor $S_\phi(k,t)$ as a function of time for the oscillator model defined by Eq.\ (\ref{eqsim}) with $\delta = \pi/4$ (non-odd coupling), $L = 1024$, and $K= 4$.}
({\it Upper panel}) $S_\phi(k,t)$ as a function of $k$. ({\it Lower panel})  $k ^{2\alpha+1} S_\phi(k,t)$ as a function of $y = k t^{1/z}$.
In all rescalings and power-law visual guides, $\alpha = 1.07$, $\alpha_s = 1.40$, $z=1.36$, and $\alpha_\text{loc}=0.97$. The structure factor curves are shown for the same time points displayed in the lower panel of Fig.~\ref{figWGdelta0250K4}. Averages based on 2000 realizations.}
\label{figSkdelta0250K4}
\end{figure}

Results for the KPZ equation with columnar disorder indicate \cite{szendro} the occurrence of faceted anomalous scaling whose full assessment can be best provided by studying correlations in Fourier space. Thus, the upper panel of Fig.~\ref{figSkdelta0250K4} shows the structure factor $S_\phi({\bf k},t)$ as a function of $k$ for different times, also enabling qualitative comparison with the results discussed in the previous section. The structure factor certainly does not follow the FV form, Eq.\  (\ref{SkFV}), but indeed shows a very good agreement with the faceted scaling form of Eq.~(\ref{Sk}). This is explicitly confirmed in the lower panel, where
the $S(k,t)$ data are collapsed closely following the expected behavior, Eq.\ (\ref{Sk}),
except for wave numbers that are close to the ultraviolet (discretization) cutoff.
Similar results ---supporting the existence of faceted scaling close to that found for the KPZ equation with columnar noise \cite{szendro}--- in the synchronization of a 1D system of oscillators with non-odd coupling are obtained for other values of $\delta \in (0,\pi/2)$, which are provided through $G_{\phi}(\ell,t)$ and $S_\phi(k,t)$ in Appendix \ref{AppB}, as well as for other coupling strengths $K>K_c$, such as $K = 3, 10$, or $20$ (not shown). One last point worth mentioning is that the study of localization effects in the multiplicative diffusion equation and directed polymers in random media, related with the KPZ equation with columnar noise, indicates that the precise time behavior of the correlation length is sub-ballistic, with $\xi(t) \sim t \ln^{-3/4} t$ \cite{nattermann, krughh}. The value we are considering here for the dynamical exponent $z$ is thus an effective one which in general is not universal (it may depend, for instance, on the noise distribution \cite{krughh}), as it facilitates the comparison with results obtained for $\delta = 0$. Moreover, this approach was shown to provide a satisfactory description in the numerical study of the KPZ equation with columnar noise, where it was compared with the sub-ballistic form just quoted \cite{szendro}.



\subsection{Dynamics of synchronization}

\begin{figure}[h!]
\hspace{-0.2cm}\includegraphics[scale=0.35]{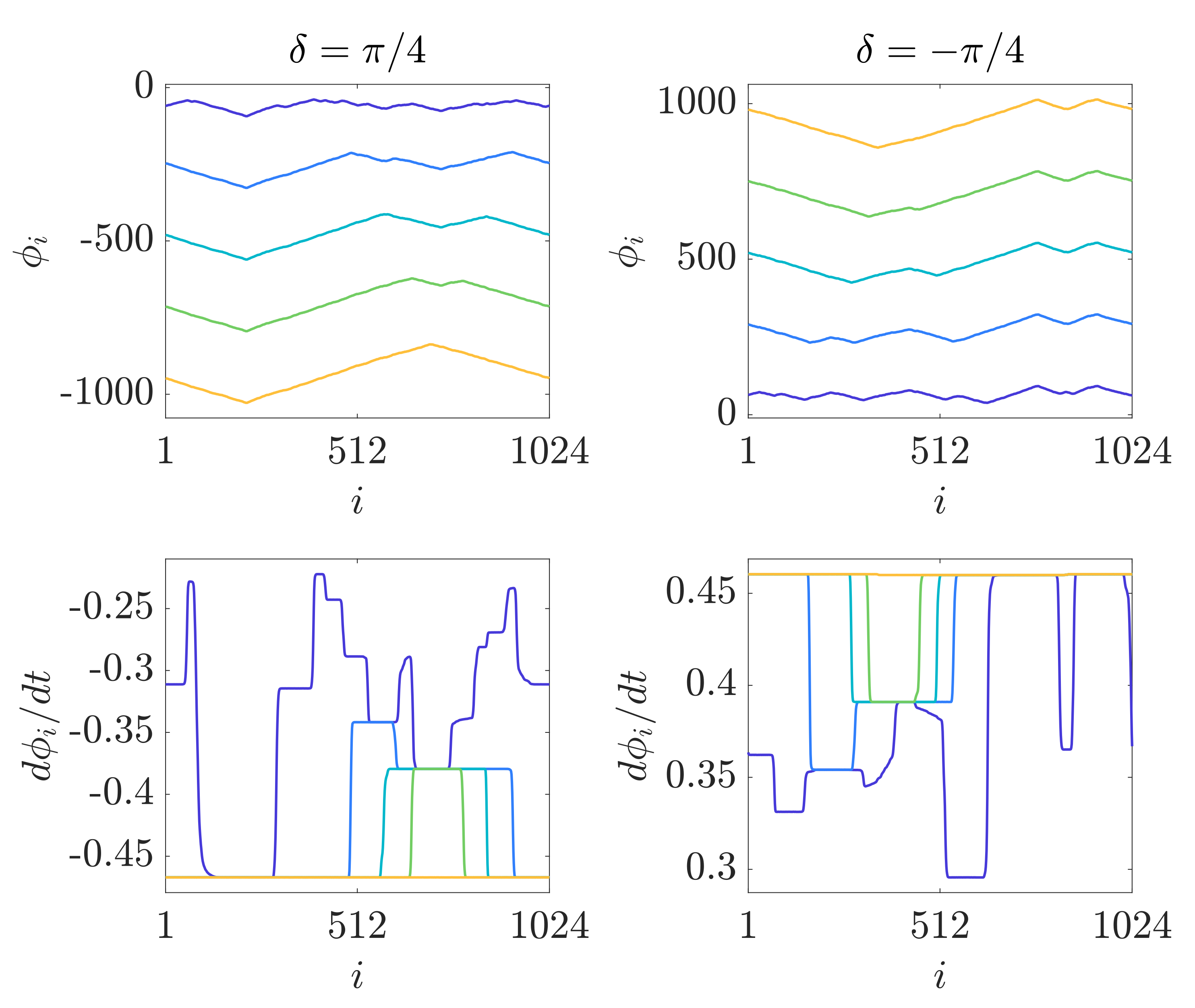}
\vspace{-0.7cm}
\caption{{\sf \bf Phases $\phi_i$ and instantaneous frequencies $d\phi_i/d t$  as functions of time in a representative trajectory of the oscillator lattice defined by Eq.\ (\ref{eqsim}), with size $L = 1024$ and coupling strength $K= 4$, for $\delta = \pi/4$ (left column) and $\delta = -\pi/4$ (right column).} ({\it Upper row}) Phases $\phi_i$ for $t = 200,700,1200, 1700$, and $2200$. Earlier times are shown in darker colors, later times in lighter colors. ({\it Lower row}) Instantaneous frequencies $d{\phi}_i/dt$ for the same trajectories, at the same time points.}
\label{FigTrajdelta025K4}
\end{figure}

In the upper panels of Fig.~\ref{FigTrajdelta025K4} we show phases $\phi_i$, and in the lower panels we show instantaneous frequencies $d \phi_i/ dt$, from representative trajectories for $\delta = \pi/4$ (left column) and $\delta = -\pi/4$ (right column). Files S2 and S3 within the Supplemental Material \cite{suppl} are animated movies for the corresponding time evolutions. In Fig.\ \ref{FigTrajdelta025K4}, different colors correspond to different time values, darker colors indicating earlier times and lighter colors later times. The oscillators rapidly form clusters that evolve at the same frequency, which is negative for $\delta = \pi/4$ and positive for $\delta = -\pi/4$, as expected from the previous discussions, see Sec.\ \ref{prelim}. These clusters merge into large ones progressively, the faster evolving one at the interface between two clusters always absorbing the slower one. Eventually, the fastest cluster of oscillators recruits more and more oscillators, some of which have been previously absorbed multiple times into progressively faster clusters, until it spans the whole system, which moves uniformly.

There is an obvious similarity with the phenomenology reported in Ref.~\cite{moroney}, for a model which (except for the addition of a uniform frequency) corresponds to $\delta = -\pi/4$, and, less evidently, to that of the KPZ equation with columnar noise \cite{szendro}. In both references the authors resort to the Cole-Hopf mapping into a diffusion problem with random traps and sources for an explanation. In the present context of systems of oscillators (also that of Ref.~\cite{moroney}), further light can be shed by means of the simple two-oscillator problem, Eq.\ (\ref{twoosc}), and its stable equilibrium point, Eq.\ (\ref{twoosccond}). Indeed, according to that model when two oscillators synchronize for $\delta \neq 0$, the effective frequency is changed by a factor proportional to $-K \sin \delta$ [see Eq.~(\ref{omegaeff2osc})], which is of course negative for $\delta = \pi/4$ and positive $\delta = -\pi/4$. Having undergone some previous coarsening, when two oscillators at the boundary of two clusters interact, their intrinsic frequencies must be ``renormalized'' (in some sense) in order to take into account the effect of the rest of the synchronized oscillators in the cluster. The result at different stages of the coarsening process, however, is always similar, in the sense that it leads to a change in the effective frequencies of the clusters under consideration that consistently pushes towards smaller/larger values, depending on whether $\sin \delta$ is positive or negative. Indeed, the space average of Eq.\ (\ref{eqsim}) yields
\begin{align}
\overline{\frac{d \phi_i}{d t}} &= \overline{\omega_i }+  K[ \overline{\sin(\phi_{i+1}\!-\!\phi_i\!+\!\delta)\!+\!\sin(\phi_i\!-\!\phi_{i+1}\!+\!\delta)} - 2 \sin \delta]\nonumber \\
&= \overline{\omega_i } - 2 K \sin \delta \left(1-\overline{\cos \Delta \phi} \right).
\label{avfreq}
\end{align}
where the average instantaneous frequency vanishes, $\overline{\omega_i} \approx \langle \omega \rangle = 0$, by the law of large numbers. Thus, $\overline{d \phi_i/d t} \approx - 2 K \sin \delta \left(1-\overline{\cos \Delta \phi} \right).$ Unless the phases are all equal, the average instantaneous frequency will be nonzero and proportional to $-K \sin \delta$. In fact, the larger the slopes $\Delta \phi$ (i.e., the smaller $\overline{\cos \Delta \phi}$ is), the larger the effective frequency is. Since the spatial averaging may be restricted to a cluster of oscillators, this may explain the fact that the steepest slopes are associated with the faster (in absolute value) clusters in Fig.~\ref{FigTrajdelta025K4}. The relationship between this and the bimodal distributions in Fig.~\ref{figslopes} in Appendix \ref{App0}, and also the behavior of the slopes in the inset of the upper panel in Fig.~\ref{figWGdelta0250K4}, is further discussed in Appendix \ref{AppA}.

From the point of view of the continuum approximation, the crucial difference between the synchronization observed for the Kuramoto model ($\delta = 0$) and for $|\delta| \in (0,\pi/2)$, 
is the existence for the latter of a KPZ nonlinearity in Eq.~(\ref{eq4}), whose coupling precisely equals $\lambda=-2Ka^2\sin\delta$, according to Eq.~(\ref{gammaders}). Such a term is well known to introduce an excess velocity that drives the interface dynamics \cite{barabasi,krug97}, which is no longer up-down symmetric. In analogy with Eq.\ \eqref{avfreq}, if one performs the space average of Eq.~(\ref{eq4}), due to the periodic boundary conditions and to the fact that the mean intrinsinc frequency is zero, the excess velocity is the integral of the KPZ nonlinearity, and its sign is therefore that of $\lambda$, indeed proportional to $-\sin \delta$.

\subsection{Fluctuation statistics}

To complete the description of the kinetic roughening universality class obtained for $\delta\neq 0$, Fig.~\ref{FigDistalldeltas} shows the distribution of the fluctuations $\varphi_i$, defined as in Eq.~(\ref{fluct}), for $\delta = \pi/4$ ($K = 4$ and again we take a larger system with $L=8192$ for this purpose), as well as for other non-zero values of $\delta$ to be discussed below. The initial time $t_0 = 50$ and several values of $\Delta t$ from $50$ to $500$ have been chosen, all within the growth regime. The same distribution is shown in Fig.~\ref{FigDistalldeltas} in linear (upper panel) and in logarithmic scales (lower panel).
The histogram 
has been rescaled and shifted so that the sample mean is zero and the sample standard deviation is one. The TW distribution for the maximum eigenvalue of random matrices in the Gaussian orthogonal ensemble (GOE)
\cite{tracy09}, also normalized, is shown as a black dotted line in the figure. This distribution is the one found for the growth dynamics of the 1D KPZ universality class when using, e.g., periodic boundary conditions \cite{kriecherbauer10,takeuchi}, and is clearly very different from the Gaussian fluctuations obtained for $\delta = 0$. Very good agreement is thus found between the numerical results for the oscillator lattice and the GOE-TW PDF. While points in the right tail appear to deviate for probabilities $\lesssim 10^{-3}$, which might be related to finite size effects, the left tail continues to follow very closely the theoretical curve for the lowest values inspected, on the order of $10^{-6}$, below which large statistical uncertainties exist (not shown). An alternative visualization of the results in Fig.~\ref{FigDistalldeltas}, including the comparison of the numerical histograms with other distribution in the TW family of PDFs is provided in Appendix \ref{AppFluct}. To our knowledge, this is the first time that an explicit connection between synchronization and the ubiquitous TW family of distributions is found. Even their relevance in the growth process given by the KPZ equation with columnar noise is not yet established, though the strong phenomenological links between that model and the synchronization model under investigation suggest that they might also characterize the fluctuations in that context.

The occurrence of the TW distribution is robust with respect to changes in $\delta$ and $K$. Indeed, when changing the sign of $\delta$,
we have found a GOE-TW with the opposite sign for the skewness, which agrees with that of $\lambda\propto-\sin\delta$, as is also the case in the standard KPZ universality class \cite{kriecherbauer10,takeuchi}. Note that, for representation purposes, the PDF curves shown
for $\delta>0$ in Fig.\ \ref{FigDistalldeltas} have been normalized to have positive skewness.
More importantly, the results reported in Fig.~\ref{FigDistalldeltas} for $\delta = \pi/8$ and $3\pi/8$ (two values of $\delta$ for which results analogous to those of Figs.~\ref{figWGdelta0250K4} and \ref{figSkdelta0250K4} for $\delta = \pi/4$ are displayed in Appendix \ref{AppB}), confirm that practically the same distribution is found for other nonzero values of $\delta \in (-\pi/2,\pi/2)$.
As for the effect of the coupling strength $K$, we also find the same kind of fluctuation PDF, which nicely follows the GOE-TW form, for smaller $K$ and for larger $K$, including values as large as $K=20$. Moreover, for all these different values of $\delta$ and $K$ the choice of the time window (as determined by $t_0$ and $\Delta t$) does not seem to be particularly important [as long as it is kept within the growth time interval where $W_\phi(t) \sim t^\beta$], nor do small modifications in the value of $\beta$ affect it. Taken together, these results suggest that the TW fluctuations are a generic and robust feature of the oscillator lattice, Eq.\ (\ref{eqsim}), for $\delta = (-\pi/2,\pi/2)$, provided that $\delta \neq 0$, at least for sufficiently large coupling strengths $K>K_c$.

The TW fluctuation PDF is currently considered as one of the universal traits induced by the KPZ  nonlinearity \cite{kriecherbauer10,takeuchi}, although examples are known of systems which, while displaying it, do not have e.g.\ the same scaling exponent values of the KPZ universality class, see e.g.\ Ref.\ \cite{marcos22} and references therein. For systems within the KPZ universality class proper, it is accompanied by a covariance that scales as Eq.~(\ref{covscal}), where the scaling function $\mathcal{C}$ is the covariance of the Airy$_1$ process \cite{bornemann} if the boundary conditions are periodic as in our present work \cite{alves,oliveira,takeuchi}. This function is precisely the one employed in the data collapse of Fig.~\ref{FigCovariances}, as explained at the end of the previous section, and is shown there as a black dotted line. But despite the TW form of the fluctuation distribution, the scaling function $\mathcal{C}(x)$ of the phase covariance $C_\phi(\ell,t)$ for $\delta = \pi/8, \pi/4$, and $3\pi/8$ seems to be identical to that observed for the Kuramoto coupling ($\delta = 0$), for which the fluctuations are Gaussian. Actually, the numerically-obtained $C_\phi(\ell,t)$ for all the values of $\delta$ that have been inspected follow quite closely the covariance of the Larkin model in Eq.~(\ref{eq:cLark}), as can be observed in Fig.~\ref{FigCovariances}, see black solid line. At this point, we note that the covariance of the EW equation with time-dependent noise is also Airy$_1$ \cite{carrasco}, as for the nonlinear KPZ equation, in spite of the fact that EW features Gaussian fluctuations.

Taken together, the results reported in this section, in combination with those of Appendices \ref{AppFluct} and \ref{AppB}, suggest that a faceted anomalous scaling and KPZ (TW) fluctuations are generic properties of synchronization for values of $\delta \neq 0$ in $(-\pi/2,\pi/2)$. The same can be said about the general appearance of the trajectories, which are for $\delta = \pi/8$ and $3\pi/8$ (not shown) qualitatively very similar to those displayed for $\delta = \pi/4$ in Fig.~\ref{FigTrajdelta025K4} and to those of the KPZ equation with columnar noise \cite{szendro}. The super-rough scaling and Gaussian fluctuations found for Kuramoto coupling ($\delta =0$) are thus rather peculiar, and likely due to the up-down symmetry (and the resulting absence of the KPZ nonlinearity in the effective continuum description) that only holds for that parameter choice. It is intriguing, though not completely unexpected for the reasons mentioned above, that this sharp distinction is not present in the phase covariance $C_\phi(\ell,t)$.

\section{Discussion and conclusions}

Our results demonstrate a very strong connection between synchronization in lattices of oscillators and kinetic roughening in systems with columnar noise. Both the dynamics of synchronizing oscillators and that of growing interfaces display GSI: provided the system is in the synchronized phase, its critical behavior does not depend on the values of the parameters $K$ and $\delta$ as long as they are generic. For $\delta=0$ the universality class differs from that of $\delta\neq 0$, but does not depend on $K$ either, provided synchronization takes place. The scaling behavior we find in all cases is anomalous, with forms which are here studied (to the best of our knowledge) for the first time in a synchronization context. While the scaling of synchronization with Kuramoto coupling $\Gamma(\Delta \phi) = \sin(\Delta \phi)$ ($\delta = 0$) is super-rough \cite{dassarma}, for other forms of Kuramoto-Sakaguchi coupling ($\delta \neq 0$), the scaling is generically faceted \cite{ramasco}, requiring one additional exponent for its description. Moreover the fluctuations, which in the former case are simply Gaussian, in the latter cases follow a TW distribution, an important characteristic of the KPZ universality class, known to be displayed by an increasing variety of low-dimensional strongly-correlated classical and quantum systems, see e.g. Ref.~\cite{makey20}. From this point of view, the behavior observed for the Kuramoto coupling ($\delta = 0$) is a singular exception to what seems to be the general rule for $0<|\delta| <\pi/2$, as has been sometimes discussed in the synchronization literature, see Ref.~\cite{ostborn} and references therein. This is most likely related to the same up-down symmetry (and corresponding lack of excess velocity) that distinguishes the EW equation from the KPZ equation in kinetic roughening systems with time-dependent noise \cite{barabasi,krug97}.

The most conspicuous phenomenological aspects of such faceted scaling of synchronization, as illustrated by the trajectories displayed in Fig.~\ref{FigTrajdelta025K4}, which present obvious similarities with the growth dynamics of the KPZ equation with columnar noise \cite{szendro}, were already apparent in a model of phase oscillators that corresponds to ours for $\delta = -\pi/4$ except for the constant term proportional to $\sin \delta$ in Eq.\ (\ref{eqsim}) that guarantees that $\Gamma(0) = 0$ \cite{moroney}. It is remarkable that such a model arises from the effective description of a system of driven-dissipative bosons, which raises the possibility of experimentally observing faceted anomalous scaling in quantum systems.

This connection between synchronization and kinetic roughening is established through a continuum effective description that relies on a perturbative expansion, which is expected to work for high enough $K>K_c$ if $\cos \delta > 0$. The resulting continuum equations in 1D are those of the Larkin model (EW equation with columnar noise) for Kuramoto coupling, and the KPZ equation with columnar noise generically for the other types of couplings investigated. Presumably the effective description is valid in higher dimension, which might explain the obvious similarities between the phenomenology of 2D synchronization in the particular instance of the model discussed above \cite{moroney} and the 2D KPZ equation with columnar noise \cite{szendro}. In this regard, note that due to the excellent agreement between the Kuramoto model and the analytical predictions of the Larkin model, the conclusion that the former is in the universality class of the EW equation with columnar disorder can hardly be questioned, and may also hold in higher dimensions [notice that the exponents in those cases can also be read from Eq.~(\ref{SkLarkin})]. For $\delta\neq0$, the type of scaling Ansatz and the scaling exponent values, as well as the sign of the skewness of the fluctuation PDF with respect to that of $\delta$, all strongly suggest the occurrence of the universality class of the KPZ equation with columnar disorder. The slight numerical disagreements that persist between the numerical values we obtain for the exponents for some $\delta$ and those reported from simulations of the KPZ equation with this type of disorder \cite{szendro} may be expected due to known properties of the latter \cite{krug97}. For example, for bounded probability distributions of the columnar disorder that depend on a parameter, corrections to scaling ensue which depend on that parameter \cite{krughh}, thus manifesting (weak) non-universal behavior. Moreover, for $d\leq 2$ the nonlinear fixed point for the KPZ equation with columnar disorder is unstable with respect to infinitely many nonlinear terms generated under the renormalization group (RG) \cite{nattermann}, possibly inducing the logarithmic corrections to the power-law growth of the correlation length mentioned in Sec.\ \ref{sec:exp_nonodd}.

Within the continuum approximation, the basic difference between the Kuramoto sine ($\delta = 0$) and the other forms of coupling ($\delta \neq 0$) lies in the KPZ nonlinearity, absent for the Kuramoto coupling due to its odd symmetry. The KPZ nonlinearity is known to cause growth along the local surface normal direction, which gives rise to a nonzero excess velocity \cite{barabasi,krug97}. Its inclusion breaks the up-down symmetry of the Larkin model, and, in the discrete lattices of oscillators under investigation, changes the (still anomalous) scaling from super-rough to faceted, and the fluctuations from Gaussian to TW distributed. Additional systems are also known in which fluctuations are TW \cite{Nicoli2013,marcos22} due to the occurrence of a KPZ nonlinearity \cite{abraham,nicoli}, even if scaling exponent values differ from those of the KPZ universality class, and irrespective of its $\lambda$ coupling being time-dependent or not. Conversely, recent examples are also available, like the tensionless KPZ equation, in which the KPZ nonlinearity induces a non-symmetric distribution \cite{cartes,rodriguez-fernandez22} which however does not take the precise TW form. Apparently, no kinetic-roughening system is yet known with TW-distributed fluctuations, but in which the KPZ nonlinearity is absent. Completing the description of the universality class, for our oscillator lattices we obtain that, surprisingly, the covariances follow the analytical form of the Larkin model regardless of the value of $\delta$. As was mentioned above, an analogous behavior is known for time-dependent noise, whereby the covariance of the linear EW and the nonlinear KPZ equations are both Airy$_1$ \cite{carrasco}. The precise reason remains unclear, although it may be related with the fluctuation-dissipation theorem accidentally satisfied by the KPZ equation in one dimension \cite{barabasi,krug97}, which makes it share a number of properties with its (EW) linear approximation.
At any rate, to our knowledge $\delta\neq0$ oscillator lattices thus provide the first known examples of TW fluctuation statistics without Airy covariance.

Related with the dynamic scaling Ansatz, recent work has shown that the Larkin model also loses its super-rough scaling in favor of faceted scaling in the presence of anharmonicities \cite{purrello}. In the synchronization context, our work raises the question whether the singularity of the Kuramoto coupling as an isolated  point displaying super-rough scaling in a parameter space where faceted scaling seems to be generic extends to other kinds of symmetry-breaking perturbations. In the kinetic roughening context, on the other hand, it raises the question whether KPZ fluctuations are also present in those other models displaying faceted scaling, which then should be added to the currently increasing list of systems known to display a TW PDF. Moreover, it would be interesting to know whether those models also display covariances like those of the linear theory (which is the Larkin model in our case), or they follow those of the Airy$_1$ process.

Overall, our work opens up new vistas on the problem of synchronization in finite dimensions by making use of the powerful conceptual framework of nonequilibrium kinetic roughening to an unprecedented level, incorporating also quite recent developments in that field. This allows us to clarify the dynamical process by which synchronous motion is achieved, which  remained a relatively poorly explored aspect of synchronization, and is here shown to be characterized by forms of nonequilibrium criticality previously observed in growth processes. Many important issues remain to be studied at the interface between these two topics of contemporary statistical physics and nonlinear science. For example, the role of thermal noise in the effective dynamics of synchronized motion, as in fact variations of the Kuramoto model and similar models of phase oscillators are frequently studied in the presence of time-dependent noise sources \cite{acebron}. The connection to the (standard, thermal noise) KPZ equation in such models has already been discussed and recently exploited in the study of routes out of synchronization \cite{lauter}, but every aspect of GSI in that context, including the nature of scaling and its fluctuations, apparently remains unknown. Finally, while we have considered synchronous dynamics for coupling strengths $K>K_c$, which guarantee that synchronization is achieved for long times, an obviously relevant goal would be to understand the transition to synchronization at the level of the effective continuum description. Ideally, this should provide accurate estimates of $K_c$, and shed new light on the role played by the symmetry of the coupling function \cite{ostborn,strogatz}.

\begin{acknowledgments}
This work has been partially supported by Ministerio de Ciencia e Innovaci\'on (Spain), by Agencia Estatal de Investigaci\'on (AEI, Spain, 10.13039/501100011033), and by European Regional Development Fund (ERDF, A way of making Europe) through Grants No.\ PGC2018-094763-B-I00 and No.\ PID2021-123969NB-I00, and by Comunidad de Madrid (Spain) under the Multiannual Agreement with UC3M in the line of Excellence of University Professors (EPUC3M23), in the context of the V Plan Regional de Investigaci\'on Cient\'{\i}fica e Innovaci\'on Tecnol\'ogica (PRICIT).

\end{acknowledgments}

\appendix

\section{Stationary slopes $\Delta \phi$ and validity of the small-slope approximation}
\label{App0}

\begin{figure*}[t!]
\hspace{-0.2cm}\includegraphics[scale=0.375]{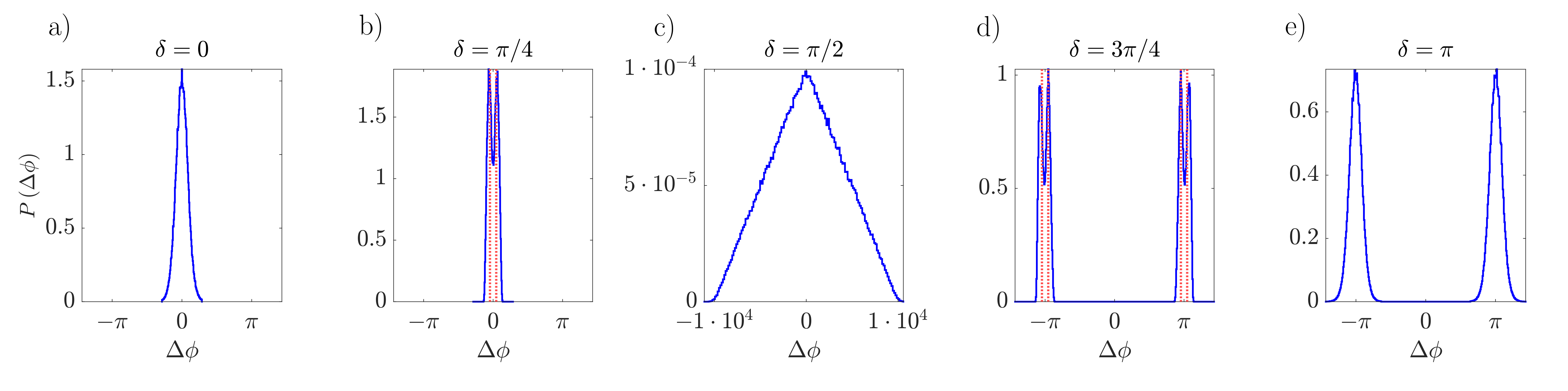}
\vspace{-0.2cm}
\caption{{\sf \bf Distributions of slopes $\Delta \phi$ for the oscillator lattice defined by Eq.\ (\ref{eqsim}) with $L=256$ and $K = 10$, for different values of $\delta$ in each panel :} (a) $\delta = 0$, (b) $\delta = \pi/4$, (c) $\delta = \pi/2$, (d) $\delta = 3\pi/4$, and (e) $\delta = \pi$. The red dotted lines in (b) and (d) indicate the position of the upper bounds for $\Delta \phi$ of the two-oscillator condition, Eq.\  (\ref{twoosccond}). Each histogram is based on at least $2000$ realizations at time $t = 5000$ (sufficiently large for synchronization to be attained).}
\label{figslopes}
\end{figure*}

To determine the range of $\delta$ for which the small-slope approximation in  Eq.~(\ref{eq4}) is expected to hold, we focus on the stationary slopes $\Delta \phi$ upon saturation in the synchronous regime. Fig.~\ref{figslopes} shows the distribution of slopes $\Delta \phi$ at long times in systems of $L=256$ oscillators for the same five values of $\delta$ displayed in Fig.~\ref{figsat}. The coupling strength of choice, $K=10$, is high enough to achieve synchronization in all cases, except (of course) for $\delta = \pi/2$.

Let us focus initially on the two bimodal distributions in panels (b) and (d), corresponding to $\delta = \pi/4$ and $3\pi/4$ (non-odd coupling), respectively. The first of the two conditions in Eq.~(\ref{twoosccond}) rightly predicts the centering of the peaks around $0$ or $\pi$, depending on the sign of $\cos \delta$. The second condition in Eq.~\eqref{twoosccond} yields the upper and lower bounds $\pm \arcsin(K\cos \delta)^{-1}$, for $\Delta \omega = \pm 2$, which are shown as red dotted lines in Fig.~\ref{figslopes} and roughly correspond to the peaks observed around $\Delta \phi = 0$ for $\delta = \pi/4$ [panel (b)] and around $\Delta \phi = \pi$ for $\delta = 3 \pi/4$ [panel (d)]. Indeed these peaks get closer to $0$ or $\pi$ (as the case may be) as $K$ is increased (not shown).

Furthermore, the two-oscillator model, Eq.\ (\ref{twoosc}), also captures qualitative aspects of the phenomenology presented in Sec.~\ref{nonodd}, some of which are briefly discussed in the original reference \cite{sakaguchi}. We summarize here those that shed light on the results reported in Fig.~\ref{figslopes}. When two oscillators lock, the frequency they achieve for $\delta\neq 0,\pi$ and $K\geq K_c^\text{SSK}$ is shifted with respect to the average frequency, see Eq.~(\ref{omegaeff2osc}), a shift that in turn must influence oscillators in some neighborhood, which need to adjust to it. The net result, as show in Fig. \ref{FigTrajdelta025K4}, is a coarsening dynamics of clusters of contiguous synchronized oscillators, which merge and, as a result, gradually become less in number and faster. Similar results were recently reported in Ref.~\cite{moroney} and, in the context of growing interfaces, Ref.~\cite{szendro}. In a sense, at each boundary between two clusters there is approximately a new two-oscillator problem, only with ``renormalized'' intrinsic frequencies that also include the effect of further oscillators. The tendency towards having a bimodal distribution of slopes can be inferred from the numerical results in Sec.~\ref{nonodd} and in Appendix \ref{AppA}, which show a gradual disappearance of slower clusters with smaller stationary slopes. The random assignment of intrinsic frequencies $\omega$ causes the finite width of the peaks.

In Figs.~\ref{figslopes}(a) and \ref{figslopes}(e) we observe unimodal distributions centered around $0$ (for $\delta = 0$) or $\pm \pi$ ($\delta = \pi$). The first condition in Eq.~(\ref{twoosccond}) once again rightly predicts the centering of the peaks. For such values of $\delta$ (odd coupling), in the two-oscillator model synchronization is achieved at the average frequency $\bar{\omega}$, see Eq.~(\ref{omegaeff2osc}). Extrapolating once more this picture to the many-oscillator problem, when two oscillators lock, the effect must propagate to neighboring oscillators, smoothening the differences in effective frequencies, which gradually become closer to the mean of the frequency distribution (here it is zero). As there are all types of (``renormalized'') effective $\Delta \omega$ in the many-oscillator problem, which moreover become smaller in absolute value with time, the unimodal distribution around zero is expected, given the proportionality of $\sin \Delta \phi$ and $\Delta \omega$ in Eq.~(\ref{twoosccond}). This situation is also analyzed in Appendix \ref{AppA} for $\delta = 0$.

 The only non-synchronous evolution  in Fig.~\ref{figslopes} is observed for $\delta = \pi/2$, panel (c), for which the slopes keep evolving at arbitrarily long times. We find a triangular distribution of $\Delta \phi$, which is known \cite{papoulis} to be the PDF of a random variable (the local slope $\Delta \phi$) defined as the difference between two uniformly-distributed variables (the two phases whose difference yields $\Delta \phi$). This is in agreement with the main conclusions drawn in Sec.~\ref{prelim} about this highly peculiar case, and in fact confirms that the overall behavior is similar to that of uncoupled oscillators, as shown in Fig.~\ref{figsat} (c).

  \begin{figure}[h!]
\includegraphics[scale=0.60]{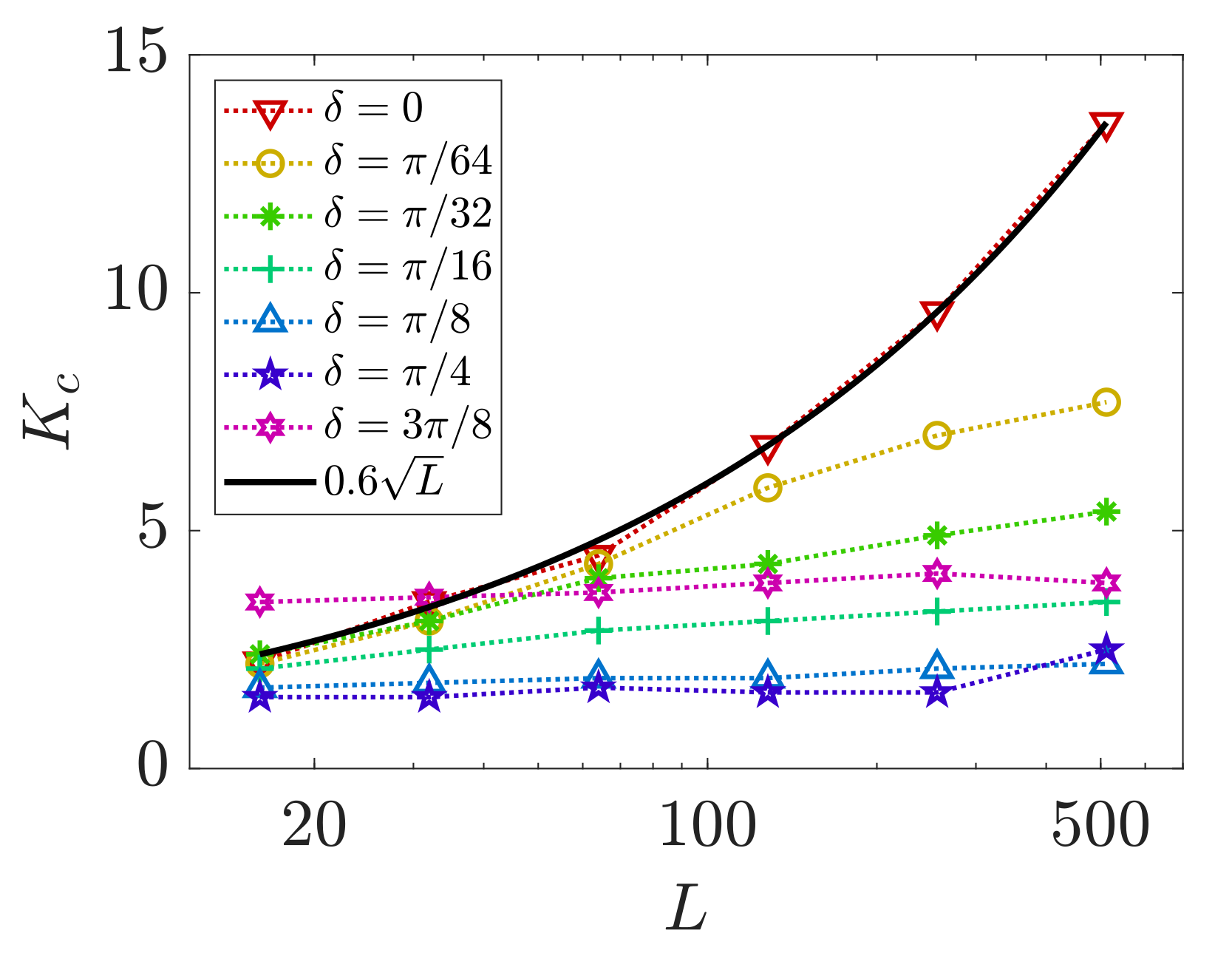}
\vspace{-0.2cm}
\caption{{\sf \bf Critical coupling strength $K_c$ as a function of $L$ for different values of $\delta$.}  Numerical estimates of $K_c$ based on the saturation of the roughness $W_{\phi}(L,t)$ (see text for details) are represented for different values of $\delta$ with various symbols and colors (see legend). The black solid line corresponds to $0.6\sqrt{L}$ growth, and is shown for comparison with the numerical estimates for $\delta = 0$.}
\label{figappa}
\end{figure}

 We next reassess the previous results from the perspective of the continuum description Eq.\ (\ref{eq4}), with a special emphasis on the validity of the small-slope approximation. This will help us restrict the range of $\delta$, and, guided by the analysis of Sec.\ \ref{kc} and Appendix \ref{AppDep}, choose appropriate values for the coupling strength $K$. We shall consider separately situations for which $\cos \delta > 0$ and those for which $\cos \delta < 0$; the $\cos \delta = 0$ condition is only to be considered as a limiting case of the former as $\delta \to \pm \pi/2$. We always assume that $K> K_c$.

When $\cos \delta > 0$, for sufficiently large $K$ the phase differences, as illustrated in Figs.~\ref{figslopes}(a) and \ref{figslopes}(b), are close to zero and the small $\Delta \phi$ approximation leading to Eq.\ (\ref{eq4}), with parameters given in Eq.~(\ref{gammaders}), is justified. As previously explained, for $\delta = 0$ the KPZ nonlinearity drops out from Eq.\ (\ref{eq4}), namely $\lambda = 0$. The lack of the corresponding lateral-growth mechanism, which would result in a non-zero average excess velocity \cite{barabasi}, sets this most-studied case apart from the others. On the other hand, as $\delta$ approaches $\pi/2$ and $-\pi/2$ ($\cos \delta \to 0$), $\nu \to 0$, and the smoothening effect of the surface tension thus disappears \cite{barabasi}. This might provide a complementary explanation for the lack of synchronization observed in panel (c) of Figs.~\ref{figsat} and \ref{figslopes}, at least for the initial stages starting from the flat initial condition, as later on the small $\Delta \phi$ approximation is expected to fail. Incidentally, for the case of time-dependent noise, the KPZ equation with zero surface tension has been recently shown to define a universality class with a scaling Ansatz that is intrinsically anomalous \cite{rodriguez-fernandez22}.

For $\cos \delta < 0$ and sufficiently large $K$, on the other hand, as shown in Figs.~\ref{figslopes}(d) and \ref{figslopes}(e), the phase differences stabilize around $\pm \pi$. That may explain why the roughness $W_{\phi}(t)$ for $\delta = 3\pi/4$ and $\pi$ in Fig.~\ref{figsat} is initially larger for $K>0$ than for the uncoupled $K=0$ case: there are `forces' pushing neighboring phases to distance from each other as much as possible.  In those cases, the perturbative expansion leading to Eq.\ (\ref{eq4}) is unjustified.

\section{Dependence of the critical coupling strength $K_c$ on the system size $L$}
\label{AppDep}

\begin{figure*}[t!]
\includegraphics[scale=0.4]{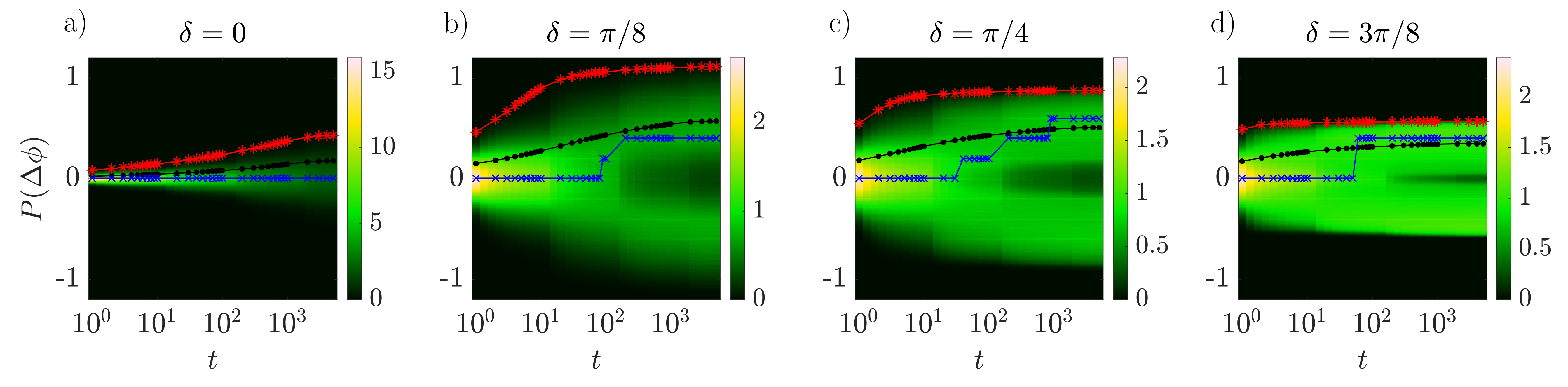}
\vspace{-0.6cm}
\caption{{\sf \bf PDF of the local slopes $P(\Delta \phi)$ across time in a system of $L=1024$ oscillators for different values of $\delta$, together with the square root of three slope observables considered in the main text.} (a) $\delta = 0$ and $K=40$, (b) $\delta = \pi/8$ and $K=4$, (c) $\delta = \pi/4$ and $K=4$, (d) $\delta = 3\pi/8$ and $K=8$. $P(\Delta \phi)$ is shown across time using the color code specified in the color bars, together with the square root of the average squared slope (black dots), that of the mean maximum squared local slope (red asterisks), and that of the the most probable value of the squared slope (blue crosses). Histograms and averages are based on 2000 realizations.}
\label{FigHists}
\end{figure*}

The critical coupling strength $K_c$ is defined as the minimum value of the coupling strength $K$ for which saturation of the roughness $W_{\phi}(L,t)$ is achieved, which implies the equality of the effective frequencies in the system. In our numerical estimation of $K_c$, we consider a reference time $t_\text{ref} = 5000$, several times larger than needed to leave the growth regime behind, and saturation is considered to take place whenever the roughness, based on $200$ disorder realizations, varies across the time window $[t_\text{ref}, 2 t_\text{ref}]$ less than 1\% of its variation across
$[0,t_\text{ref}]$, i.e.\!
\begin{equation}
\left|\frac{W_{\phi}(L,2 t_\text{ref}) - W_{\phi}(L,t_\text{ref})}{W_{\phi}(L,t_\text{ref}) - W_{\phi}(L,0)}\right| < 0.01 .
\end{equation}
Strictly speaking, this is expected to provide only a lower bound for $K_c$, as the inclusion of further realizations might show that a larger value of $K$ is needed to achieve synchronization for all possible disorder configurations.

In Fig.\ \ref{figappa} we show the critical coupling strength $K_c$ thus obtained as a function of the system size $L$ for various values of $\delta$, including $\delta = 0, \pi/8,\pi/4$, and $3\pi/8$, which were already shown in Fig.\ \ref{figPD} in a different representation. As explained in Sec.~\ref{kc}, while for $\delta = 0$ we have $K_c\propto \sqrt{L}$, for those other values of $\delta$ the dependence on $L$ is much weaker. In fact, there is a consistent moderate increase for the largest sizes inspected which could indicate a mild dependence on $L$, or simply reflect that the statistical threshold implied in the numerical determination of $K_c$ becomes comparatively easier to be exceeded for large system sizes.

Some values of $\delta$ not considered elsewhere in our analysis, namely $\delta = \pi/16, \pi/32$, and $\pi/64$, are included in Fig.\ \ref{figappa} so as to provide some insight into the behavior of $K_c$ as $\delta \to 0$. One can see that $K_c$ follows the dependence observed for $\delta = 0$ for small sizes $L$, beyond which there is a gradual change towards $\delta\neq 0$ behavior. Such crossover gets displaced towards larger values of $L$ as $\delta$ becomes smaller. In the effective continuum approximation, this might be related to the minimal length scale needed for the effect of the KPZ nonlinearity to be observable at very small values of $\lambda\propto \sin \delta$.

\section{Time-dependent distribution of slopes $\Delta \phi$}
\label{AppA}

While discussing the insets in the upper panels of Figs.~\ref{figWGdelta0K40} and \ref{figWGdelta0250K4} (see also Figs.~\ref{figWGdelta0125K4} and \ref{figWGdelta0375K8} in Appendix \ref{AppB}) various observables related to the time evolution of the distribution of the slopes $\Delta \phi$ were mentioned. Specifically, there we discussed the evolution of the average squared slope $\langle\overline{(\Delta \phi)^2} \rangle$, as well as the maximum squared local slope (defined as the square of the greatest slope in the system, averaged over realizations) and the most probable squared local slope (defined as the squared local slope that has a higher probability in the distribution). A more complete picture emerges by inspecting the distribution of local slopes $P(\Delta \phi)$ itself across time, which is what we show in this Appendix. In fact we have already shown some instances of $P(\Delta \phi)$ in Fig.~\ref{figslopes}, but they correspond to single snapshots of the slope distribution for long times. Moreover, as they were used to restrict the range of parameters based on the validity of the small-slope approximation, they included values of $\delta$ that were finally not studied in Secs.~\ref{odd} and \ref{nonodd}. Here we focus on $\delta \in [0,\pi/2)$ for the same values of $K$ chosen in the main text and in Appendix \ref{AppB} ($K\approx 2 K_c$). For $\delta \in (-\pi/2,0)$, we obtain equivalent results due to the symmetry discussed at the end of Sec.\ \ref{nummod}.

\begin{figure*}[t!]
\includegraphics[scale=0.37]{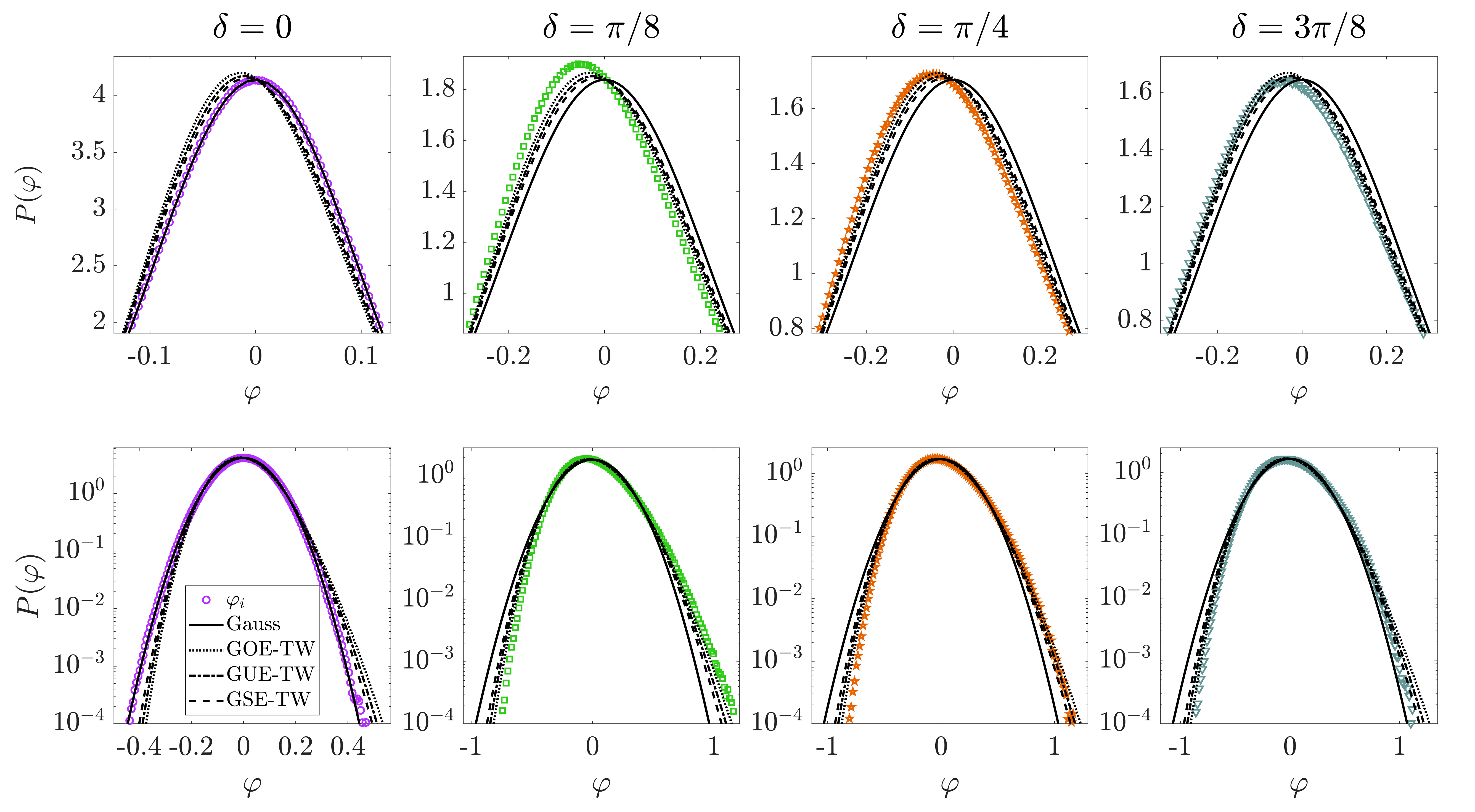}
\vspace{-0.3cm}
\caption{{\sf \bf Histograms of fluctuations $\varphi_i$ defined as in Eq.\ \eqref{fluct} for the oscillator lattice, Eq.\ \eqref{eqsim}, in a system of $L=8192$ oscillators, for $\delta = 0$ ($K=110$), $\delta = \pi/8$ ($K=4$), $\delta = \pi/4$ ($K=4$), and $\delta = 3 \pi/8$ ($K=8$).} Each column corresponds to a value of $\delta$, as indicated above the upper row panels. ({\it Upper row}) Linear scale. ({\it Lower row}) Logarithmic scale. We always take as reference time $t_0 = 50$, and values of $\Delta t = 50, 100, 150, \ldots, 500$, which are in the interval where a power-law growth $W_\phi(t) \sim t^\beta$ is observed. The black solid line corresponds to a Gaussian distribution, the dotted line to a GOE-TW distribution, the dash-dotted line to a GUE-TW distribution and the dashed line to a GSE-TW distribution, each of them being normalized to the sample mean and variance of the corresponding histogram, based on $10^4$ realizations.}
\label{FigAppendixFluct}
\end{figure*}

In Fig.~\ref{FigHists} we show the PDF of the slopes $P(\Delta \phi)$ for four values of $\delta$. Those are the values that were fully discussed in the main text, namely $\delta = 0$ [panel (a), which was considered in Sec.~\ref{odd}] and $\delta = \pi/4$ [panel (c), which was considered in Sec.~\ref{nonodd}], as well as two values, $\delta = \pi/8$ [panel (b)] and $\delta = 3\pi/8$ [panel (d)], for which only the fluctuation PDF and the covariances were studied in the main text (see Figs.~\ref{FigDistalldeltas} and \ref{FigCovariances}, respectively, and the pertinent discussion in Sec.~\ref{nonodd}), while the scaling Ansatz and exponent values are addressed in Appendix \ref{AppB}. The PDF of the slopes $P(\Delta \phi)$ is represented as it evolves across time using the colors specified in the color bars. Additionally, we include the observables that we showed in the insets of the upper panels of Figs.~\ref{figWGdelta0K40} and \ref{figWGdelta0250K4} (and also  Figs.~\ref{figWGdelta0125K4} and \ref{figWGdelta0375K8} in Appendix \ref{AppB}). In fact, as they are dimensionally squared slopes, we take their square root to make the comparison with $P(\Delta \phi)$ possible. Thus, we show the square root of the average squared slope (black dots), that of the mean maximum squared local slope (red asterisks) and that of the the most probable value of the squared slope (blue crosses).


For $\delta = 0$, panel (a), $P(\Delta \phi)$  is unimodal, and the most probable value is always close to $\Delta \phi =0$. The distribution simply spreads over a larger interval of $\Delta \phi$ as time increases, until it saturates. For the remaining values of $\delta$, panels (b), (c), and (d), the distribution starts being unimodal close to the flat initial condition, but at some point it becomes bimodal, and the most probable value becomes distinctly different from zero. This seems to be something that Eq.~(\ref{avfreq}) can explain, as it shows that for a typical cluster of oscillators (or a cluster of oscillators which is large enough) such that the average of the involved intrinsic frequencies approaches zero, the absolute values of the average instantaneous frequencies of the cluster approach $2 K \sin \delta \left(1-\overline{\cos \Delta \phi} \right)$, where $\overline{\cos \Delta \phi}$ is the spatial average of $\cos \Delta \phi$ over the cluster. Therefore a faster cluster requires larger slopes (smaller values of $\cos \Delta \phi$), and when the system starts being dominated by a few fast clusters (see Fig.~\ref{FigTrajdelta025K4}) there is a reshuffling of the slopes towards larger (absolute) values. These three cases, $\delta = \pi/8$, $\pi/4$, and $3\pi/8$, correspond to an effective description given by the KPZ equation with columnar noise, with faceted anomalous scaling and KPZ fluctuations (see Sec.~\ref{nonodd} and Appendix \ref{AppB}), so their qualitative similarity is expected. The most conspicuous effect of modifying $\delta(\neq 0)$ is the following: As $\delta$ increases, the spread of the distribution $P(\Delta \phi)$ decreases and the saturation of the largest values occurs earlier.

\section{Fluctuation PDFs}
\label{AppFluct}

In this appendix we provide an alternative visualization of the fluctuation PDFs included in Fig.\ \ref{FigDistalldeltas}. Specifically, in Fig.\ \ref{FigAppendixFluct} we show the histograms of the fluctuations defined as in Eq.\ (\ref{fluct}), but without normalizing them to zero mean and unit variance. Panels in different columns correspond to different values of $\delta$ (specifically, $0, \pi/8, \pi/4$, and $3\pi/8$), for the same parameter values used in Fig.\ \ref{FigDistalldeltas}, with representations in linear (upper row) and logarithmic (lower row) scales. The theoretical curves are adjusted to the sample mean and variance of the histograms in each case. They correspond to a Gaussian PDF and to the three different types of TW PDF giving the distribution of the largest eigenvalue of random Hermitian matrices \cite{kriecherbauer10,takeuchi,tracy09,Fortin2015}, namely, the Gaussian orthogonal ensemble (GOE), the Gaussian unitary ensemble (GUE), and the Gaussian symplectic ensemble (GSE). The numerical histograms for $\delta \neq 0$ follow quite closely the GOE-TW PDF (already considered in the main text, and in Fig.\ \ref{FigDistalldeltas}) across a wide range of values, more so than they do for any of the other two TW PDFs, as shown by systematic effects at the peak and in the tails of the distribution. This is in line with many results in the kinetic-roughening literature showing that 1D systems with periodic boundary conditions displaying TW fluctuations are described by the GOE-TW PDF, see e.g.\ Refs.\  \cite{kriecherbauer10,takeuchi} and references therein. For $\delta = 0$, however, the fluctuation PDF is clearly Gaussian.

\begin{figure}[h!]
\includegraphics[scale=0.47]{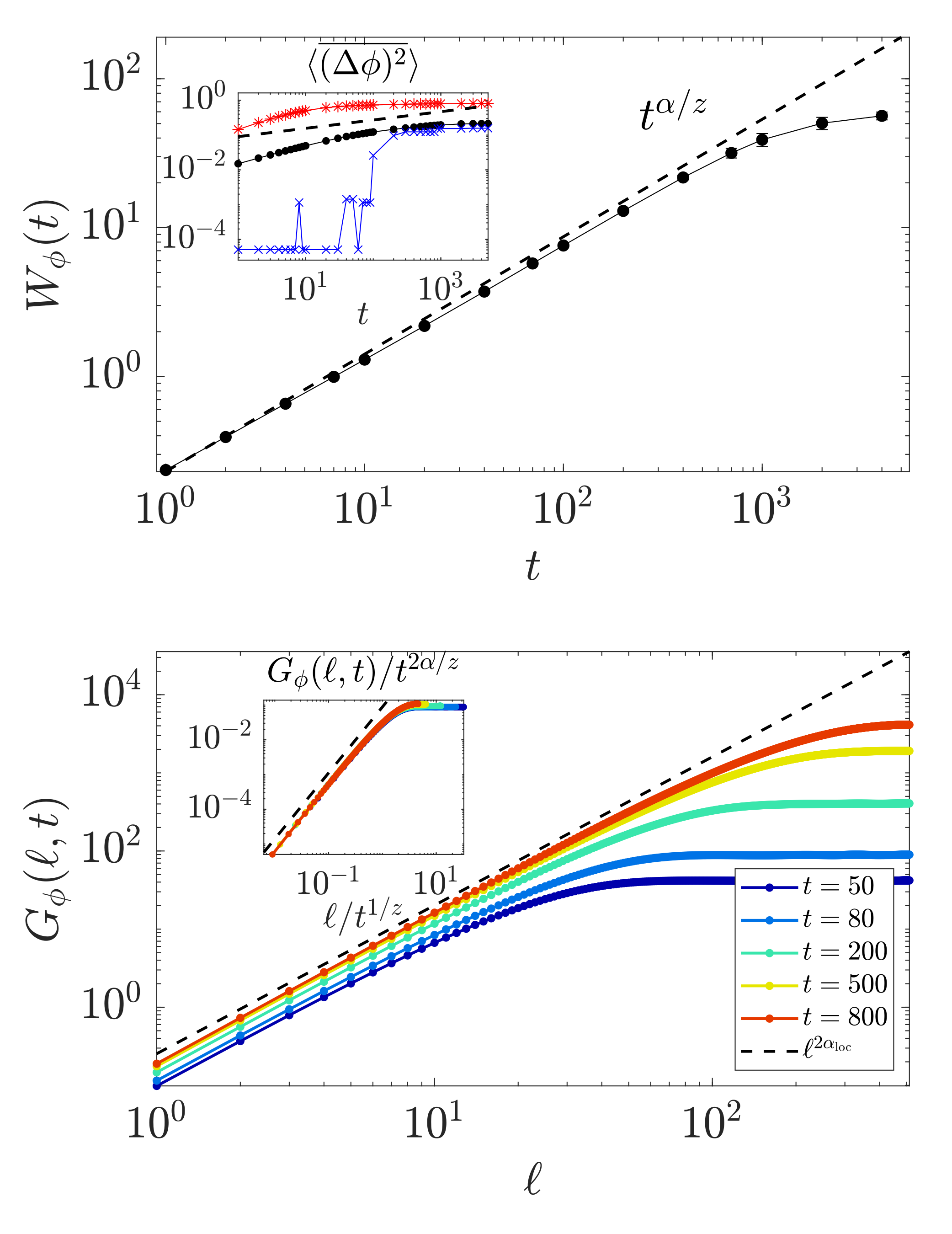}
\vspace{-0.3cm}
\caption{{\sf \bf Roughness $W_\phi(t)$ and height-difference correlation function $G_\phi(\ell,t)$ for the oscillator lattice, Eq.\  (\ref{eqsim}), with $\delta = \pi/8$, $L = 1024$, and $K= 4$.}
({\it Upper panel}) $W_\phi(t)$ as a function of time in the synchronous regime, with error bars displaying the standard error of the average across realizations. In the power-law visual guides we have used $\alpha = 1.12$ and $z=1.42$. The inset shows the average squared slope $\langle\overline{(\Delta \phi)^2} \rangle$ (black circles), as well as the average of the maximum squared local slope (red asterisks), and the most probable value of the local slope (blue crosses), as a function of time for the same parameter values (see text for definitions). The black dashed line shows the power-law growth of the average squared slopes predicted by the theory, $t^{2(\alpha-\alpha_\text{loc})/z}$ for $\alpha_\text{loc}=0.95$. ({\it Lower panel}) $G(\ell,t)$ as a function of $\ell$ in the synchronous regime for different times (see legend). The dashed line shows the power law growth with the distance $l^{2\alpha_\text{loc}}$. Inset: Rescaling of $G(\ell,t)$ following the theoretical form in Eq.~(\ref{glt_mod}). Averages based on 2000 realizations.
}\label{figWGdelta0125K4}
\end{figure}

\begin{figure}[h!]
\includegraphics[scale=0.47]{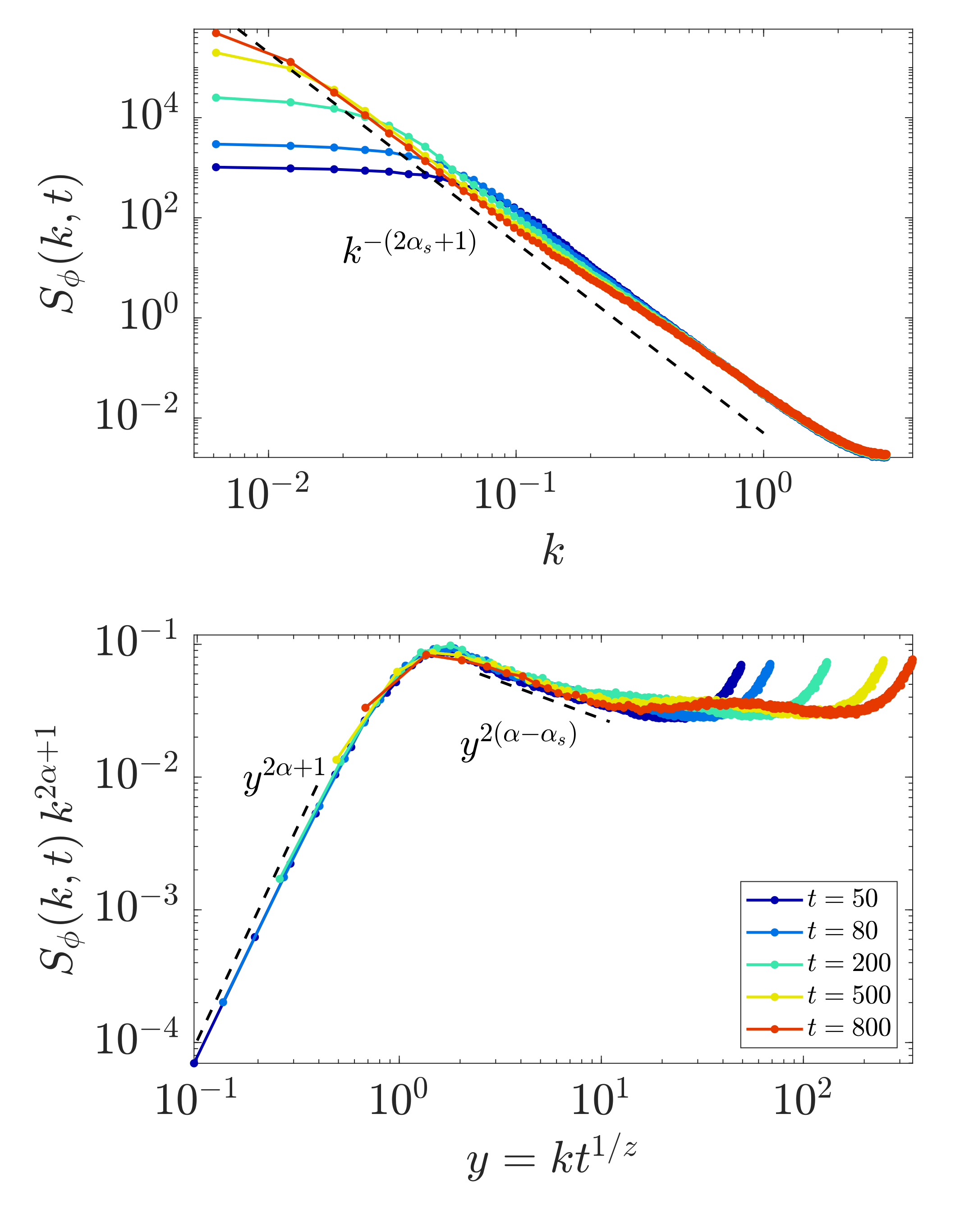}
\vspace{-0.3cm}
\caption{{\sf \bf Structure factor $S_\phi(k,t)$ as a function of time for the oscillator lattice, Eq.\ (\ref{eqsim}), with $\delta = \pi/8$, $L = 1024$, and $K= 4$.} ({\it Upper panel}) $S_\phi(k,t)$ as a function of $k$. ({\it Lower panel}) $k ^{2\alpha+1} S_\phi(k,t)$ as a function of $y = k t^{1/z}$ In all rescalings and power-law visual guides, $\alpha = 1.12$. $\alpha_s = 1.40$, $z=1.42$, and $\alpha_\text{loc}=0.95$. The structure factor curves are shown for the same time points displayed in the lower panel of Fig.~\ref{figWGdelta0125K4}. Averages based on 2000 realizations.}
\label{figSkdelta0125K4}
\end{figure}

\section{Scaling Ansatz and exponents for other values of $\delta$}
\label{AppB}

In the following we show results analogous to those displayed in Sec.~\ref{nonodd} for $\delta = \pi/4$, but for other values of $\delta \in (0,\pi/2)$. Specifically, we study the scaling of the synchronization dynamics for $\delta = \pi/8$ and $3\pi/8$, which were considered in the analysis of the fluctuations and the covariances, see Figs.~\ref{FigDistalldeltas} and \ref{FigCovariances}. As all the elements in the figures were previously discussed for $\delta = \pi/4$ in connection with Figs.~\ref{figWGdelta0250K4} and \ref{figSkdelta0250K4} and the qualitative conclusions are the same, the descriptions will be brief.

In Fig.~\ref{figWGdelta0125K4} we show the roughness $W_\phi(t)$ (upper panel) and height-difference correlation function $G_\phi(\ell,t)$ (lower panel) for $\delta = \pi/8$ in a system of $L=1024$ oscillators. The critical coupling $K=4$ is roughly twice as large as the critical coupling strength obtained from the saturation of the width as described at the end of Sec.\ \ref{satsta} and in Appendix \ref{AppDep}. The exponents values are $\alpha = 1.12$, $\alpha_s = 1.40$, $z=1.42$, and $\alpha_\text{loc}=0.95$, with uncertainties again below $0.05$ for the same reasons discussed for $\delta = \pi/4$ in the main text, which differs slightly in the exponent values. The inset of the upper panel shows again the mean, maximum, and most probable value of the squared slopes averaged across realizations, and that of the lower panel shows a rescaling of the two-point correlations based on the theory of anomalous scaling, Eq.\ (\ref{glt_mod}). All the comments given for $\delta = \pi/4$ apply in this case too.

\begin{figure}[h!]
\includegraphics[scale=0.47]{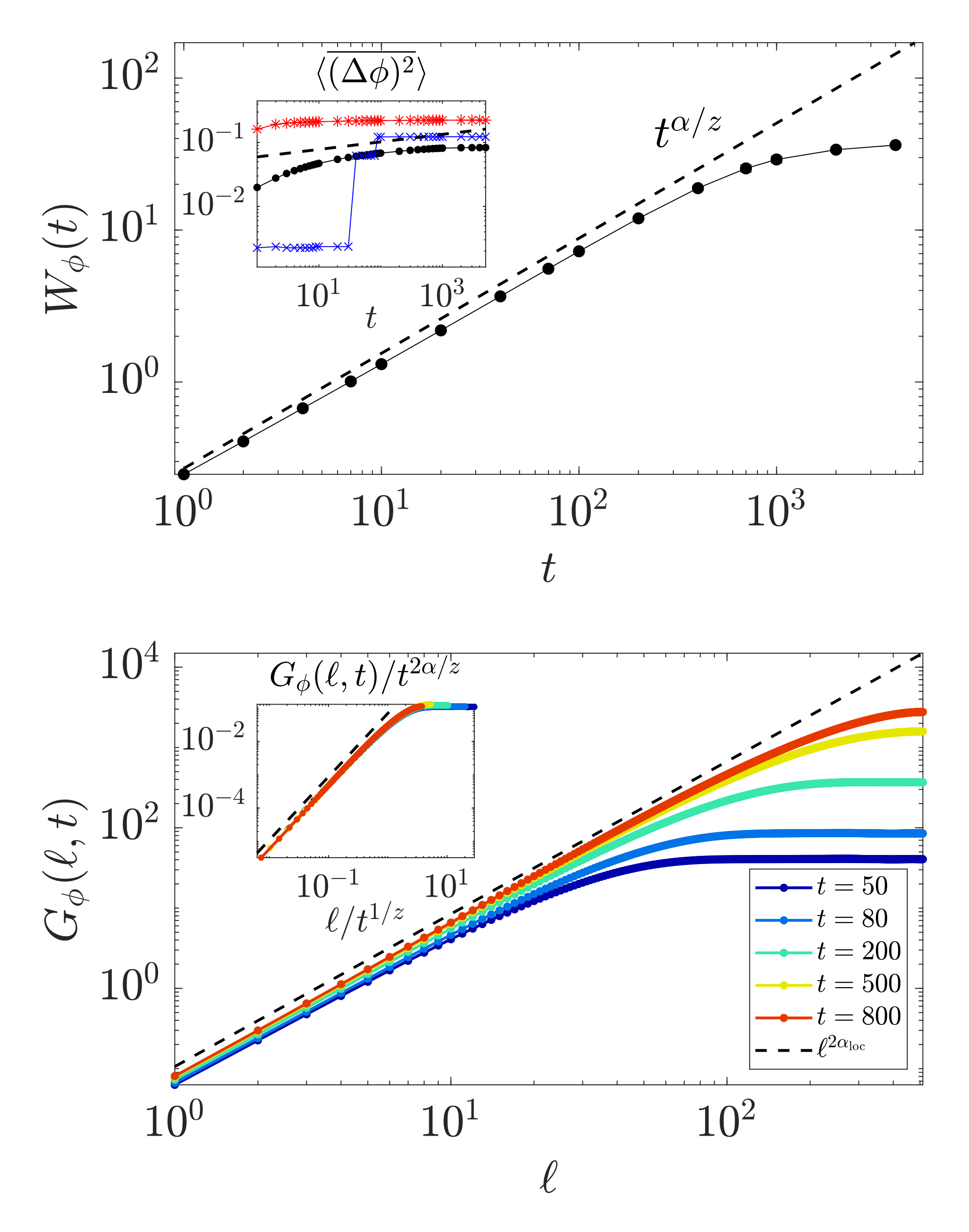}
\vspace{-0.3cm}
\caption{{\sf \bf Roughness $W_\phi(t)$ and height-difference correlation function $G_\phi(\ell,t)$ for the oscillator lattice, Eq.\ (\ref{eqsim}), with $\delta = 3\pi/8$, $L = 1024$, and coupling strength $K= 8$.}
({\it Upper panel}) $W_\phi(t)$ as a function of time in the synchronous regime, with error bars displaying the standard error of the average across realizations. In the power-law visual guides we have used $\alpha = 1.03$ and $z=1.36$. The inset shows the average squared slope $\langle\overline{(\Delta \phi)^2} \rangle$ (black circles), as well as the average of the maximum squared local slope (red asterisks), and the most probable value of the local slope (blue crosses), as a function of time for the same parameter values (see text for definitions). The black dashed line shows the power-law growth of the average squared slopes predicted by the theory, $t^{2(\alpha-\alpha_\text{loc})/z}$ for $\alpha_\text{loc}=0.95$. ({\it Lower  panel}) $G(\ell,t)$  as a function of $\ell$ in the synchronous regime for different times (see legend). The dashed line shows the power law growth with the distance $l^{2\alpha_\text{loc}}$. Inset: Rescaling of $G(\ell,t)$ following the theoretical form in Eq.~(\ref{glt_mod}). Averages based on 2000 realizations.
}\label{figWGdelta0375K8}
\end{figure}

The structure factor $S_\phi(k,t)$ also for $\delta = \pi/8$ (with the same parameter choices and exponents) is shown in its original form (upper panel) and in rescaled form (lower panel) in Fig.~\ref{figSkdelta0125K4}. As in the case of $\delta = \pi/4$, the scaling is anomalous of the faceted type discussed in Ref.~\cite{ramasco}.

We next focus on analogous results for $\delta = 3\pi/8$. In Fig.~\ref{figWGdelta0375K8} we show the roughness $W_\phi(t)$ (upper panel) and height-difference correlation function $G_\phi(\ell,t)$ (lower panel) in a system of $L=1024$ oscillators. The critical coupling $K=8$ is again roughly twice as large as the critical coupling strength obtained from the saturation of the width. The exponents of choice are $\alpha = 1.03$, $\alpha_s = 1.40$, $z=1.36$, and  $\alpha_\text{loc}=0.95$, with uncertainties again below $0.05$, for the same reasons discussed for $\delta = \pi/4$ in the main text, with slightly different exponents. The inset of the upper panel shows again the mean, maximum, and most probable value of the squared slopes averaged across realizations, and that of the lower panel shows a rescaling of the two-point correlations based on the theory of anomalous scaling, Eq.\ (\ref{glt_mod}). All the comments given for $\delta = \pi/4$ apply in this case too. The most remarkable difference is perhaps the abruptness in the change of the most probable value of the squared slope (upper panel inset, blue stars), which was also shown in Fig.~\ref{FigHists} in Appendix \ref{AppA}.

\begin{figure}[h!]
\includegraphics[scale=0.47]{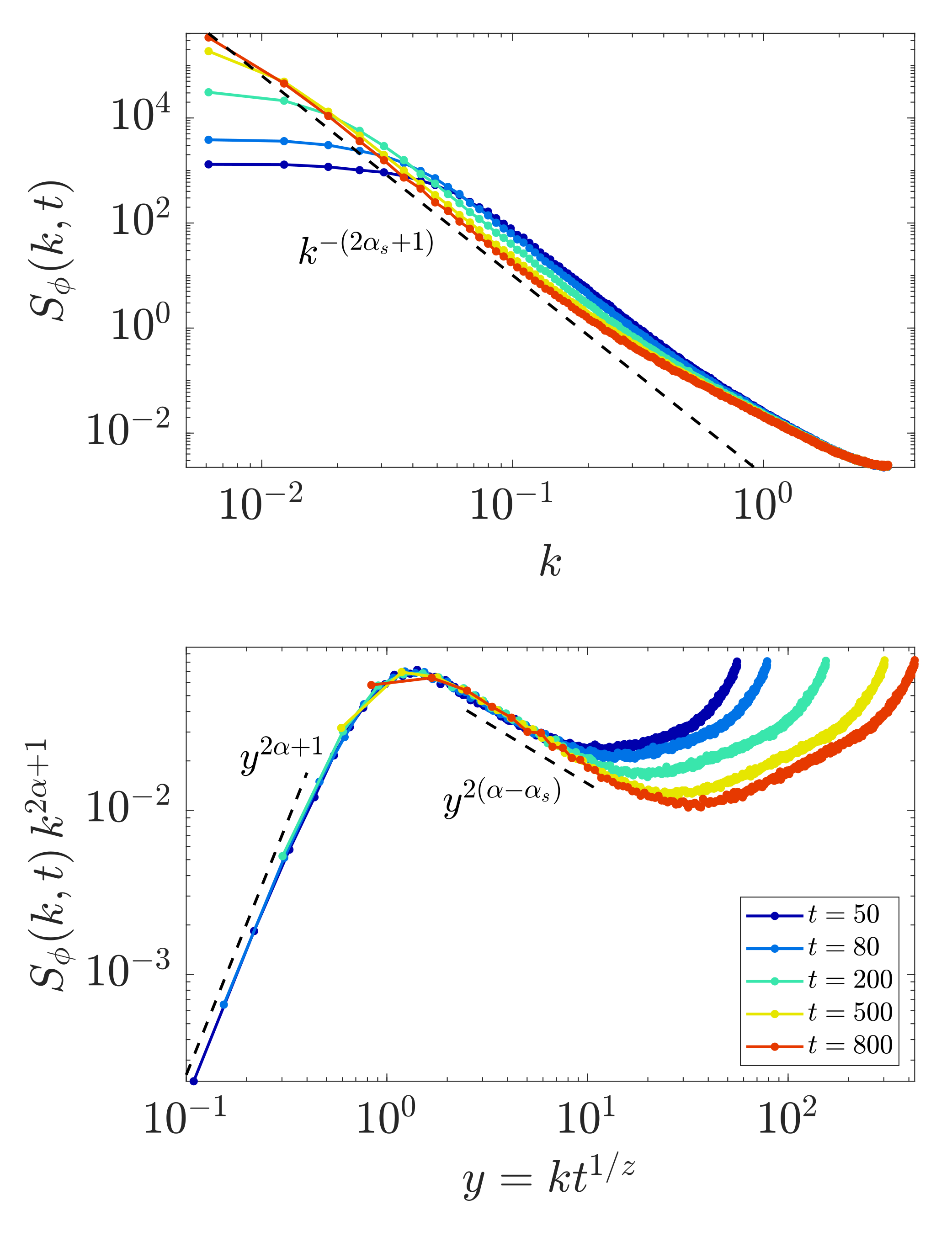}
\vspace{-0.3cm}
\caption{{\sf \bf Structure factor $S_\phi(k,t)$ as a function of time for the oscillator lattice, Eq.\  (\ref{eqsim}), with $\delta = 3\pi/8$, $L = 1024$, and $K= 8$.}
({\it Upper panel}) $S_\phi(k,t)$ as a function of $k$. ({\it Lower  panel}) $k ^{2\alpha+1} S_\phi(k,t)$ as a function of $y = k t^{1/z}$ In all rescalings and power-law visual guides, $\alpha = 1.03$, $\alpha_s = 1.40$, $z=1.36$, and $\alpha_\text{loc}=0.95$. The structure factor curves are shown for the same time points displayed in the lower panel of Fig.~\ref{figWGdelta0375K8}. Averages based on 2000 realizations.}
\label{figSkdelta0375K8}
\end{figure}

The structure factor $S_\phi(k,t)$ for $\delta = 3\pi/8$ (with the same parameter choices and exponents) is shown in its original form (upper panel) and in rescaled form (lower panel) in Fig.~\ref{figSkdelta0375K8}. As in the cases of $\delta = \pi/8$ and $\delta = \pi/4$, the scaling is anomalous of the faceted type discussed in Ref.~\cite{ramasco}.


\bibliography{syncgrowthBib}{}

\end{document}